\documentclass[a4paper,10pt]{article}
\usepackage{axodraw}
\usepackage{pstricks}
\usepackage{amsmath,amssymb,amsfonts}
\usepackage{bm}
\usepackage{epsfig}
\usepackage{color}

\catcode`\@=11

\@addtoreset{equation}{section}

\def\@normalsize{\@setsize\normalsize{15pt}\xiipt\@xiipt
\abovedisplayskip 14pt plus3pt minus3pt%
\belowdisplayskip \abovedisplayskip
\abovedisplayshortskip \z@ plus3pt%
\belowdisplayshortskip 7pt plus3.5pt minus0pt}

\def\small{\@setsize\small{13.6pt}\xipt\@xipt
\abovedisplayskip 13pt plus3pt minus3pt%
\belowdisplayskip \abovedisplayskip
\abovedisplayshortskip \z@ plus3pt%
\belowdisplayshortskip 7pt plus3.5pt minus0pt
\def\@listi{\parsep 4.5pt plus 2pt minus 1pt
     \itemsep \parsep
     \topsep 9pt plus 3pt minus 3pt}}

\relax

\catcode`@=12

\catcode`\@=11

\newcommand{\rouge}[1]{\textcolor{red}{#1}}
\newcommand{\darkgreen}[1]{\textcolor[rgb]{0.00,0.59,0.00}{#1}}
\newcommand{\bleu}[1]{\textcolor{blue}{#1}}

\def\l{\lambda}

\title{Introduction to the Pinch Technique}
\author{Vincent Mathieu
\\ \small Groupe de Physique Nucl\'{e}aire
Th\'{e}orique, Universit\'{e} de Mons-Hainaut, \\\small Acad\'{e}mie universitaire
Wallonie-Bruxelles, Place du Parc 20, BE-7000 Mons, Belgium
\\\small E-mail:
vincent.mathieu@umh.ac.be}

\begin{document}
\maketitle

\abstract{These notes are a short introduction to the pinch technique. We present the one-loop
calculations for basic QCD Green's functions. The equivalence between the pinch technique and
the background field method is explicitly shown at the one-loop level. We review the
absorptive pinch technique in the last sections. These lectures are a compilation of relevant
papers on this subject and are prepared for the third Modave Summer School in Mathematical
Physics.}

\newpage \tableofcontents

\section{Introduction}
When quantizing  gauge theories in  the continuum one must invariably resort to an appropriate
gauge-fixing procedure in order to  remove  redundant (non-dynamical) degrees of freedom
originating from the gauge invariance of the theory\footnote{The gauge-invariant formulation
of non-Abelian gauge theories on the lattice does not need ghosts or any sort of gauge-fixing,
see \cite{Wilson:1974sk}. One can also implement manifestly gauge invariant formulations of
the Exact renormalization group without gauge fixing, see \cite{mescouilles} and references
therein.}. Thus, one adds to the gauge invariant (classical) Lagrangian ${\cal L}_{\rm I}$ a
gauge-fixing term ${\cal L}_{\rm GF}$, which allows for the consistent derivation of Feynman
rules.  At this point  a new  type of redundancy makes its appearance, this time at the level
of the building blocks defining the perturbative expansion. In particular, individual
off-shell Green's functions ($n$-point functions) carry  a  great deal of unphysical
information, which disappears when physical observables  are formed. $S$-matrix elements, for
example,  are independent of the gauge-fixing scheme and  parameters chosen  to quantize  the
theory,  they are gauge-invariant (in  the sense of current conservation), they are unitary
(in  the sense of conservation of probability), and  well behaved at high energies. On  the
other hand Green's functions depend explicitly (and generally non-trivially) on the
gauge-fixing parameter entering  in the definition of  ${\cal L}_{\rm  GF}$, they grow much
faster than physical amplitudes at high energies (e.g. they grossly violate the
Froissart-Martin bound \cite{Froissart:ux}), and display  unphysical thresholds. Last but not
least, in the  context of the standard path-integral quantization by means of the
Faddeev-Popov Ansatz, Green's functions satisfy complicated Slavnov-Taylor identities
\cite{Slavnov:1972fg} involving ghost fields, instead of the usual Ward identities generally
associated  with the original gauge invariance.

The  above observations imply  that in  going from  unphysical
Green's functions to physical  amplitudes, subtle field theoretical
mechanisms are at work, implementing vast cancellations among the
various Green's functions. Interestingly, these cancellations may be
exploited in a very particular way by the Pinch Technique (PT)
\cite{Cornwall:1982zr,Cornwall:1989gv, Papavassiliou:1990zd,
Degrassi:1992ue}: A  given physical amplitude is reorganized  into
sub-amplitudes, which have the same kinematic properties as
conventional $n$-point functions (self-energies, vertices, boxes)
but, in addition, they are endowed with important physical
properties. This has been accomplished diagrammatically, at the one-
and two-loop level, by recognizing that \bleu{longitudinal momenta}
circulating  inside vertex and box diagrams generate, by
``pinching'' out internal fermion lines, propagator-like terms. The
latter are reassigned to conventional self-energy graphs in order to
give rise to effective Green's functions which manifestly reflect
the properties generally associated with physical observables. In
particular,  the  PT Green's  function  are \bleu{independent  of
the gauge-fixing  scheme} and parameters  chosen  to quantize the
theory ($\xi$ in covariant gauges, $n_\mu$ in axial gauges,  etc.),
they are gauge-invariant, {\it  i.e.}, they \bleu{satisfy} simple
tree-level-like \bleu{Ward identities} associated with the gauge
symmetry of the classical Lagrangian ${\cal L}_{\rm I}$, they
\bleu{display}  only \bleu{physical  thresholds},  and, finally,
they are \bleu{well behaved at high energies}.

Given the subtle nature of the problem, the question naturally
arises, what set of physical criteria must be satisfied by a
resummation algorithm, in order for it to qualify as ``physical''.
In other words, what are the guiding principles, which will allow
one to determine whether or not the resumed quantity carries any
physically meaningful information, and to what extend it captures
the essential underlying dynamics? To address these questions in
this notes, we postulate a set of field-theoretical requirements
that we consider crucial when attempting to define a proper
resumed propagator. Our considerations propose an answer to the
question of how to analytically continue the
Lehmann--Symanzik--Zimmermann formalism~\cite{LSZ} in the
off-shell region of Green's functions in a way which is manifestly
gauge-invariant and consistent with unitarity. In addition, we
demonstrate that the off-shell Green's functions obtained by the
pinch technique satisfy all these requirements. In fact, these
requirements are, in a way, inherent within the PT approach, as we
will see in detail in what follows.

In particular, the following is required from an off-shell,
one-particle irreducible (1PI), effective  two-point function:
\begin{itemize}

\item[(i)] \darkgreen{Resummability}. The effective two-point functions
must be resummable. For the conventionally defined two-point
functions, the resummability can be formally derived from the path
integral. In the $S$-matrix PT approach, the resummability of the
effective two-point functions is more involved and must be based
on a careful analysis of the structure of the $S$-matrix to higher
orders in perturbation theory \cite{JP&AP}.

\item[(ii)] \darkgreen{Analyticity of the off-shell Green's
function}. An analytic two-point function has the property that
its real and imaginary parts are related by a dispersion relation
(DR), up to a maximum number of two subtractions. The latter is a
necessary condition when considering renormalizable Green's
functions, as we will discuss in Sec. \ref{vm:sec:absorptivePT}.

\item[(iii)] \darkgreen{Unitarity and the optical relation}. In
the conventional framework, unitarity is defined only for on-shell
$S$-matrix elements, leading to the familiar optical theorem (OT)
for the forward scattering. Here, we postulate the validity of the
optical relation for the off-shell Green's function, when embedded
in an $S$-matrix element, in a way which will become clear in what
follows. An important consequence of this requirement is that the
imaginary part of the off-shell Green's function should not
contain any unphysical thresholds. As a counter-example, it is
shown in \cite{Papavassiliou:1996zn} that this pathology is in
fact induced by the quantum fields in the background-field-gauge
(BFG) method \cite{Dea,tH,Deb,Abb} for $\xi_Q\not= 1$.

\item[(iv)] \darkgreen{Gauge invariance}. As has been mentioned above,
one has to require that the effective Green's functions are
gauge-fixing parameter (GFP) independent and satisfy Ward identities
in compliance with the classical action. For instance, the latter is
guaranteed in the BFG method but not the former. This condition also
guarantees that gauge invariance does not get spoiled after Dyson
summation of the GFP-independent self-energies. In some of the
recent literature, the terms of gauge invariance and gauge
independence have been used for two different aspects. For example,
in the BFG the classical background fields respect gauge invariance
in the classical action. However, this fact does not ensure that the
quantum fields respect some form of quantum gauge invariance,
neither does imply that some kind of a Becchi-Rouet-Stora (BRS)
symmetry~\cite{BRS} is present for the fields inside the quantum
loops after fixing the gauge of the theory \cite{Morris,Schenk}. In
our discussion, when referring to gauge invariance, we will
encompass both meanings, {\em i.e.}, gauge invariance of the
tree-level classical particles as well as BRS invariance of the
quantum fields. A direct but non-trivial consequence of the gauge
invariance and of the abelian-type Ward identitites that the
effective off-shell Green's functions satisfy is that for large
asymptotic momenta transfers ($s\to\infty$), the self-energy under
construction must capture the running of the gauge coupling, as it
happens in Quantum ElectroDynamics (QED). Because of the
abelian-type WIs and on account of resummation, the above argument
can be generalized to $n$-point functions. In addition, the
off-shell $n$-point transition amplitudes should display the correct
high-energy limit as is dictated by the Equivalence Theorem
\cite{EqTh}.

\item[(v)] \darkgreen{Multiplicative renormalization}. Since we are
interested in renormalizable theories, {\em i.e.}, theories
containing operators of dimension no higher than four, the off-shell
Green's functions calculated within an approach should admit
renormalization. However, this requirement alone is not sufficient
when resummation is considered. The appearance of a two-point
function in the denominator of a resummed propagator makes it
unavoidable to demand that renormalization be \bleu{multiplicative}.
Otherwise, the analytic expressions will suffer from spurious
ultraviolet (UV) divergences. Particular examples of the kind are
some ghost-free gauges, such as the light-cone or planar gauge
\cite{planar}.

\item [(vi)] \darkgreen{Position of the pole}. Since the position of the
pole is the only gauge-invariant quantity that one can extract from
conventional self-energies, any acceptable resummation procedure
should give rise to effective self-energies which do not shift the
position of the pole. This requirement drastically reduces the
arbitrariness in constructing effective two-point correlation
function.

\end{itemize}

The first question one may wonder in this context is about the
conceptual and phenomenological \bleu{advantages} of being able to
work with such special Green's functions. We briefly discuss here
some concept where the pinch technique was used.

\begin{itemize}

\item
\darkgreen{QCD  effective    charge:} The unambiguous extension  of
the concept of  the gauge-independent, renormalization group
invariant, and process independent \cite{Grunberg:1992mp} effective
charge from QED to QCD \cite{Cornwall:1976ii,Cornwall:1982zr}, is of
special interest for several reasons \cite{Watson:1996fg}. The PT
construction of this quantity
 accomplishes
the explicit identification  of the conformally-(in)variant subsets
of QCD  graphs \cite{Brodsky:1982gc}, usually assumed in the field
of renormalon calculus \cite{Mueller:1992xz}.

\item
\darkgreen{Breit-Wigner resummations, resonant  transition
amplitudes, unstable particles:} The  Breit-Wigner procedure used
for regulating the physical singularity appearing in the  vicinity
of resonances ($\sqrt{s}\sim M$) is equivalent to a
\bleu{reorganization} of the perturbative series \cite{Veltman:th}.
In particular, the Dyson summation of the self-energy $\Pi(s)$
amounts to removing a particular piece from  each order of the
perturbative expansion, since from  all the Feynman graphs
contributing to a given order $n$ one only picks the part which
contains $n$ self-energy bubbles $\Pi(s)$, and  then one takes  $n
\to \infty$. Given  that non-trivial cancellations  involving the
various Green's function is generally taking place at any  given
order of the conventional perturbative  expansion, the act of
removing  one of them from each order  may distort those
cancellations, this  is indeed what happens when constructing
non-Abelian \bleu{running widths}. The pinch technique ensures that
all unphysical contributions contained inside $\Pi(s)$ have been
identified and properly discarded, \bleu{before} $\Pi(s)$ undergoes
resummation \cite{JP&AP,Papavassiliou:1995fq}.

\item
\darkgreen{Off-shell form-factors:} In non-Abelian theories, their
proper definition poses in general problems related to the gauge
invariance \cite{Fujikawa:fe}. Some representative cases have been
the magnetic dipole and electric quadrupole moments of the $W$
\cite{Papavassiliou:ex}, the top-quark magnetic moment
\cite{Papavassiliou:1993qe}, and the neutrino charge radius
\cite{Bernabeu:2000hf}. The pinch technique allows for an
unambiguous definition of such quantities. Most notably, the
gauge-independent, renormalization-group- invariant, and
target-independent neutrino charge radius constitutes a genuine
\bleu{physical} observable, since it can be extracted (at least in
principle) from experiments \cite{Bernabeu:2002nw}.

\item \darkgreen{Schwinger-Dyson  equations:}  This infinite
system of coupled non-linear integral equations  for all Green's functions of the theory is
inherently  non-perturbative  and can accommodate phenomena  such as  chiral symmetry breaking  and
dynamical mass generation.   In  practice one  is  severely limited  in their  use, and a
self-consistent  truncation scheme is needed. The main problem in this context is  that the
Schwinger-Dyson equations are built out  of gauge-dependent Green's  functions. Since the
cancellation  mechanism  is  very subtle,  involving  a  delicate conspiracy of terms  from
\bleu{all orders}, a  casual truncation often gives   rise   to gauge-dependent  approximations
for ostensibly gauge-independent quantities \cite{Cornwall:1974vz,Marciano:su}. The role of the
pinch technique in this problem is to trade the  conventional Schwinger-Dyson series for another,
written  in terms  of  the new, gauge-independent building blocks
\cite{Cornwall:1982zr,Mavromatos:1999jf,Sauli:2002tk}. The upshot of this program is then to
truncate this  new series,  by keeping only a  few terms  in a ``dressed-loop'' expansion, and
maintain exact gauge-invariance, while at the same time accommodating non-perturbative effects.
Further explanations of this non perturbative pinch technique can be found in \cite{Binosi:2007pi}.

\item
Other interesting applications include the gauge-invariant
formulation of the $\rho$ parameter at one-\cite{Degrassi:1993kn}
and two-loops \cite{Papavassiliou:1995hj}, various finite
temperature calculations \cite{Nadkarni:1988ti}, a novel approach to
the comparison of electroweak data with theory
\cite{Hagiwara:1994pw}, resonant CP violation
\cite{Pilaftsis:1997dr}, the construction of the two-loop PT quark
self-energy \cite{Binosi:2001hy}, and more recently the issue of
particle mixings \cite{Yamada:2001px}.

\end{itemize}

\bigskip

The algorithms of the S-matrix and the intrinsic pinch technique,
presented in these notes, allow us to extract Green's functions with
appropriate physical properties. These algorithms are easily perform
at the one-loop level. Nevertheless, even though the generalisation
up to an arbitrary number of loop is straightforward, the
calculation become quickly tedious. Fortunately, a correspondence
between the PT and the background field method was found and allow
us to apply directly the Feynman rules of the latter to obtain the
same Green's functions.

The Background Field Method (BFM) was first introduced by
DeWitt~\cite{Dea} as a technique for quantizing gauge field theories
while retaining explicit gauge invariance. In its original
formulation, DeWitt worked only for one-loop calculations. The
multi-loop extension of the method was given by 't Hooft~\cite{tH},
DeWitt~\cite{Deb}, Boulware~\cite{Bo}, and Abbott~\cite{Abb}. Using
these extensions of the background field method, explicit two-loop
calculations of the $\beta$ function for pure Yang-Mills theory was
made first in the Feynman gauge~\cite{Abb,IO}, and later in the
general gauge~\cite{CM}.

Both pinch technique and background field method have the same interesting feature. The
Green's functions (gluon self-energies, proper gluon-vertices, etc.) constructed by the two
methods retain the explicit gauge invariance, thus obey the naive Ward identities. As a
result, for example, a computation of the QCD $\beta$-function coefficient is much simplified.
The only thing we need to do is to construct the gauge-invariant gluon self-energy in either
method and to examine its ultraviolet-divergent part. Either method gives the same correct
answer~\cite{Cornwall:1989gv, Abb}. The connection between the background field method and the
pinch technique was first established for the one- \cite{Denner:1994nn} and two-loop
\cite{Papavassiliou:1999az} levels. The extension of the pinch technique to all orders for QCD
was carried out and the connection was shown to persist also to all orders in
\cite{Binosi:2003rr}. This connection also hold in the electroweak sector where the PT
construction to all orders was performed in \cite{Binosi:2004qe}.

\medskip

These notes are a compilation of many papers on this subject. We
refer the interested readers to these references for further
developments. We present in these lectures notes two versions of
the pinch technique for a Yang-Mills theory. The S-matrix pinch
technique and the intrinsic pinch technique are explain in
Sec.~\ref{vm:sec:pinch} on the example of the gluon self-energy.
We review the background field method in Sec.~\ref{vm:sec:BFM}
where the equivalence of the BFM and the conventional effective
actions is proved. We show also that the BFM gluon self-energy
correspond to the PT gluon self-energy at the one-loop level. In
Sec.~\ref{vm:sec:3and4gluons}, we develop the construction on the
gauge-invariant proper 3-gluon vertex by the intrinsic pinch
technique and the background field method, and we give the Ward
identity satisfied by this proper vertex. The Ward identity
satisfied by the gauge-invariant 4-gluon vertex is also presented.
Some basic concepts of quantum field and their implications are
reviewed in Sec. \ref{sec:firstprinciples}. The absorptive pinch
technique construction is presented in
Sec.~\ref{vm:sec:absorptivePT}. The first appendix is devoted to
Ward and Slanov-Taylor identities. Finally, a short development
with the dimensional regularization and the Feynman rules are
relegated to the appendix B and C.

\section{The Pinch Technique}\label{vm:sec:pinch}
\subsection{The S-matrix pinch technique}

The gluon self-energy is the simplest example that demonstrate how
the pinch technique works. We define the gluon self-energy
$i\Pi_{\mu\nu}$\footnote{A factor $i$ is added in the definition for
latter convenience.} as the sum of all one particle irreducible
diagrams with two gluon legs. The one loop contributions are
\begin{equation}
\label{vm:SEoneloop}
\begin{picture}(450,30)(0,0)
\includegraphics[scale=0.5]{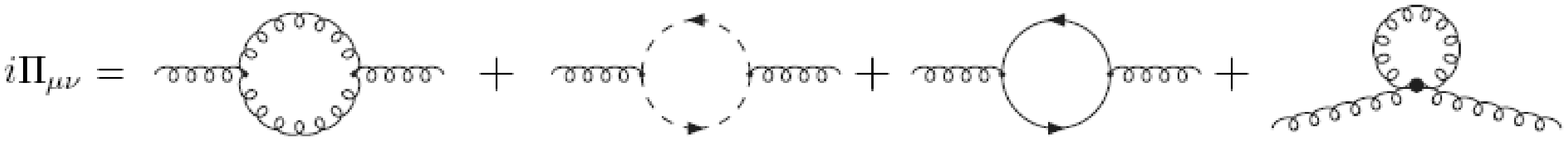}
\end{picture}
\end{equation}
This self-energy is transverse thanks to the Ward identity (see
Appendix \ref{vm:sec:WI}), like in QED, and can be written
\begin{equation}\label{}
    \Pi_{\mu\nu}(q) = t_{\mu\nu}(q)\Pi(q^2),
\end{equation}
with the transverse projector $t_{\mu\nu}(q)=g_{\mu\nu}-q_\mu q_\nu
q^{-2}$. This implies that the gluon remains massless at any order
in perturbation theory and also that $\Pi$ can be renormalized with
only one renormalization factor. A straightforward calculation shows
the this conventional (renormalized) gluon self-energy, written here
at the one-loop level with $n_f$ flavours of massless fermion
\begin{equation}\label{}
    \Pi(q^2) = \frac{g^2q^2}{16\pi^2}\left\{\left[\left(-\frac{13}{6}+\frac{\xi}{2}\right)N +
    \frac{2}{3}n_f\right]\ln(\frac{q^2}{\mu^2}) + \left(\frac{97}{36}+\frac{\xi}{2}+\frac{\xi^2}{4}\right)N
    -\frac{10}{9}n_f\right\},
\end{equation}
depends on the gauge-fixing parameter $\xi$, in the Lorentz
covariant gauges, defined by the free gluon propagator
\begin{equation}\label{vm:eq:gluon_prop}
    \Delta_{\mu\nu}^{(0)}(q) = \frac{-i}{q^2}\left[g_{\mu\nu} - (1-\xi)\frac{q_\mu
    q_\nu}{q^2}\right].
\end{equation}
Resuming all one-particle irreducible graphs, the full gluon
two-point function reads
\begin{equation}\label{}
    \Delta_{\mu\nu} =-i\left[t_{\mu\nu}(q)\Delta(q^2)+\xi\frac{q_\mu
    q_\nu}{q^4}\right],
    \quad\text{where }
    \quad \Delta(q^2) = \frac{1}{q^2-\Pi(q^2)}.
\end{equation}
Note that the dressed propagator depends on the gauge-fixing
parameter on a trivial way, given by the tree level propagator
(see Appendix \ref{vm:sec:WI}), but also through the self-energy.
Although it is possible to sum the renormalized gluon self-energy
in a Dyson series to give a radiatively corrected gluon
propagator, the quantity defined by analogy with the QED effective
charge is in general gauge-, scale- and scheme-dependent, and at
asymptotic $q^2$ does not match on to the QCD running coupling
defined from the renormalization group. The interested readers may
find more information on the gauge-invariant QCD effective charge
in \cite{Watson:1996fg}. Let us mention also that in axial gauges,
the gluon propagator is not multiplicatively
renormalizable~\cite{Cornwall:1982zr}.

Others pathologies of conventional Green's functions can be seen on
the example on the Higgs-boson self-energy
\cite{Papavassiliou:1995fq}. At the one loop level, the $W$
corrections,
\begin{equation*}
\begin{picture}(450,50)(0,0)
\includegraphics[scale=0.5]{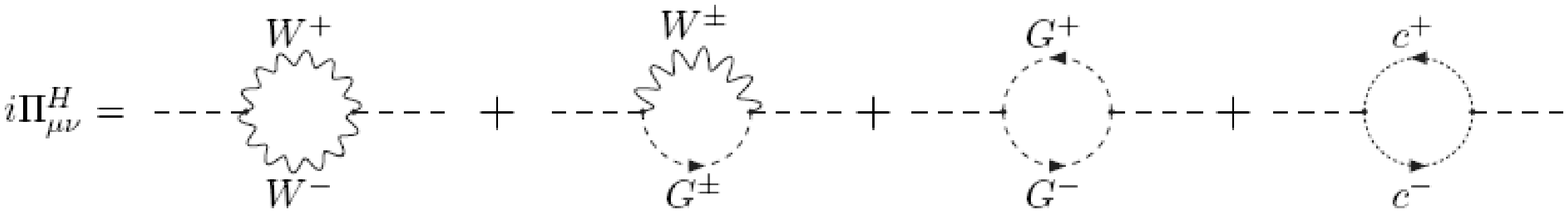}
\end{picture}
\end{equation*}
leads to a Higgs self-energy,
\begin{equation}\label{}
\begin{split}
    \Pi_{H} = \frac{\alpha_{W}}{4\pi}\bigg[&\left(\frac{(q^2)^2}{4M_W^2}-q^2
    +3M_W^2\right)B_0(q^2,M_W^2,M_W^2) \\
    &+\frac{M_H^4-(q^2)^2}{4M_W^2}B_0(q^2,\xi_W M_W^2,\xi_W
    M_W^2)\bigg],
\end{split}
\end{equation}
with
\begin{equation}\label{}
B_0(q^2,m_1^2,m_2^2)=(2\pi\mu)^{4-d}\int\frac{d^dk}{i\pi^2}\frac{1}{(k^2-m_1^2)[(k+q)^2-m_2^2]},
\end{equation}
which clearly develops unphysical threshold, i.e.
$\xi_W$-dependent threshold. Moreover the $(q^2)^2$ term violates
the Froissard-Martin bound (see Sec.\ref{sec:firstprinciples}). As
explained in \cite{Papavassiliou:1995fq}, these pathologies
disappear after the pinch technique has been carried out.

In QED the photon self-energy (also called the vacuum polarisation
tensor) is easily computed at the one loop level,
\begin{equation}\label{}
    \Pi_{\text{QED}}^{(1)}(q^2) =\frac{\alpha}{3\pi} q^2
    \left[\ln\left(\frac{q^2}{m^2}\right) - \frac{5}{3}\right],
\end{equation}
since the only diagram is a fermion loop. $m$ is the electron mass
and we used the on-shell regularization scheme. This expression
show us that, first, the photon remains massless and, secondly,
the strength of the interaction increases with the Euclidean
momentum $q^2$. The modification of the photon propagator induced
by a fermion loop can be absorbed in the definition of the fine
structure
\begin{equation}\label{}
    \alpha(q^2) = \frac{\alpha}{1-(\alpha/3\pi)
    \ln(q^2/Am^2)},
\end{equation}
with $A=\exp(5/3)$. As we will see, it is also possible to define
an effective charge gauge-independent for QCD.

\bigskip

We now review the S-matrix pinch technique as it applies to the
effective propagator. The idea is to begin with something we know to
be gauge invariant, the S-matrix, and extract from this the
corresponding gauge-invariant Green's function. Note that it is the
proper self-energy will be gauge invariant, the propagator has a
trivial gauge dependence through the free propagator and this
induces an equally trivial dependence in the two-points Green's
function.

\begin{figure}
\begin{center}
 \includegraphics[scale=0.5]{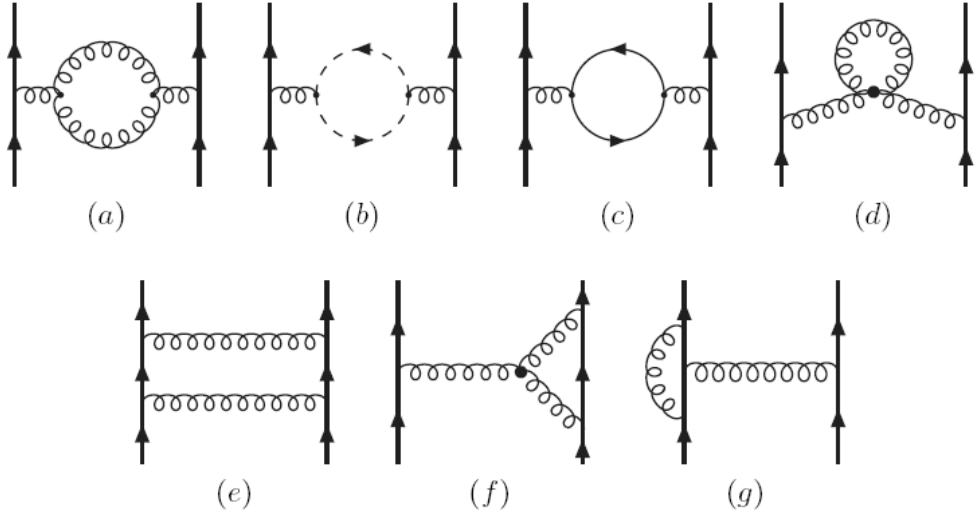}
\end{center}\caption{Fermions scattering at order $g^4$} \label{vm:fig:qqscat69}.
\end{figure}

Consider the S-matrix element T for the elastic scattering of two
fermions of masses $m_1$ and $m_2$. By asymptotic freedom,
perturbation theory is relevant for large momenta, i.e. in the
ultra-violet region, the quarks are free states and satisfy the
Dirac equation $(p\!\!\!/-m)u(p)=iS^{-1}(p)u(p)=0$. To any order in
perturbation theory, T is independent of the gauge-fixing parameter
$\xi$ \cite{Weinberg}. But, as an explicit calculation shows, the
conventionally defined proper self-energy at the one-loop level
depends on $\xi$. At the one-loop level, this dependence is
cancelled by contributions from other graphs, such as (e), (f) and
(g) in Fig.~\ref{vm:fig:qqscat69}, which, at first glance, do not
seem to be propagator-like. Note that the graphs (f) and (g) have
a mirror counterpart and the graph (e) has a crossed counterpart
not shown in Fig.~\ref{vm:fig:qqscat69}. That this cancelation
must occur and can be employed to defined a gauge-invariant
self-energy, is evident from the decomposition
\begin{equation}\label{vm:eq:defT}
    T(s,t,m_1,m_2) =
    T_1(t,\xi)+T_2(t,m_1,m_2,\xi)+T_3(s,t,m_1,m_2,\xi),
\end{equation}
where the function $T_1(t)$ depends only on the Mandelstram variable
$t=-(p'_1-p_1)^2=-q^2$, and not on $s=(p_1+p_2)^2$ or on the
external masses. Typically, self-energy, vertex, and box diagrams
contribute to $T_1$, $T_2$ and $T_3$, respectively. Moreover, such
contributions are $\xi$ dependent. However, as the sum
$T(s,t,m_1,m_2)$ is gauge invariant, it is easy to show that Eq.
\eqref{vm:eq:defT} can be recast in the form
\begin{equation}\label{vm:eq:defThat}
    T(s,t,m_1,m_2) =
    \hat T_1(t)+\hat T_2(t,m_1,m_2)+\hat T_3(s,t,m_1,m_2),
\end{equation}
where the $\hat T_i\ (i=1,2,3)$ are \bleu{separately} $\xi$
independent. To proof this assertion we derive
Eq.~\eqref{vm:eq:defT} to obtain $\partial^2T_{3}/\partial
s\partial\xi=0$. Hence $T_3$ can be written as
\begin{equation}\label{}
    T_3(s,t,m_1,m_2,\xi) = \hat
    T_3(s,t,m_1,m_2)+T'_3(t,m_1,m_2,\xi).
\end{equation}
The part $T'_3$ is added to $T_2$ and we can iterate this process to
obtain the decomposition~\eqref{vm:eq:defThat}.

\SetWidth{.5}
The propagator-like parts of graphs such as
Fig.~\ref{vm:fig:qqscat}, which enforce the gauge independence of
$T_1(t)$, are called ``pinch parts''. The pinch part emerge every
time a gluon propagator or an elementary three-gluon vertex
contribute a longitudinal momentum $k_\mu$ to the original graph's
numerator. The action of such a term is to trigger an elementary
Ward identity of the form
\begin{eqnarray}\label{vm:eq:WIelem}
 \nonumber
  k^\mu\gamma_\mu\equiv k\!\!\!/ &=& (p\!\!\!/+k\!\!\!/-m)-(p\!\!\!/-m), \\
   &=& iS^{-1}(p+k)-iS^{-1}(p),
\end{eqnarray}
once it gets contracted with a $\gamma$ matrix. The first term on
the right-hand side of Eq. \eqref{vm:eq:WIelem} will remove the
internal fermion propagator, that is a ``pinch'', whereas
$S^{-1}(p)$ vanish on shell.

\begin{figure}[tb!h]
\begin{center}
  \includegraphics[scale=0.5]{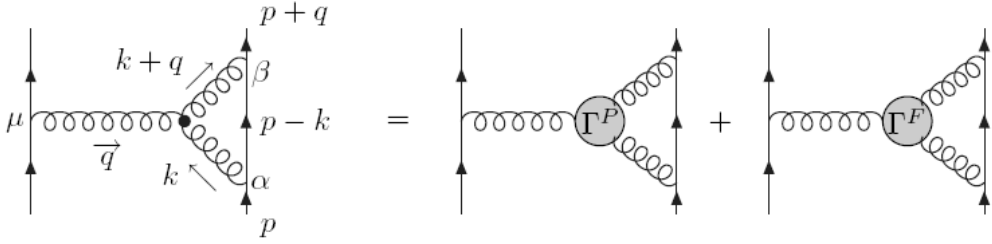}
  \caption{\label{vm:fig:vertex}Decomposition of the vertex diagram in its pinch and regular parts.}
\end{center}\end{figure}

Returning to the decomposition of Eq.\eqref{vm:eq:defT}, the
function $\hat T_1$ is gauge invariant and may be identified with
the contribution of the new propagator. We can construct the new
propagator, or equivalently $\hat T_1$, directly from the Feynman
rules. In doing so it is evident that any value for the gauge
parameter $\xi$ may be chosen, since $\hat T_1$, $\hat T_2$ and
$\hat T_3$ are all independent of $\xi$. The simplest of covariant
gauges is certainly the Feynman gauge ($\xi=1$), which removes the
longitudinal part of the gluon propagator. Therefore the only
possibility for pinching in four-fermions amplitudes arises from the
momentum of the three-gluon vertices, and the only propagator-like
contributions come from graph of Fig. \ref{vm:fig:vertex} (and its
mirror counterpart).

The amplitude of this diagram ($\bm T_a$ are the generators of the
gauge group)
\begin{equation}\label{vm:amplvertex}
    \bar u_1(ig\bm T_a )\gamma^\mu u_1
    \frac{(-i)}{q^2}gf^{abc}\int\frac{d^4k}{(2\pi)^4}
    \frac{\Gamma_{\alpha\mu\beta}(k,q)}{k^2(k+q)^2}
    \bar u_2(ig\bm T_b)\gamma^\alpha S(p-k)(ig\bm T_c)\gamma^\beta u_2
\end{equation}
has a pinch part $i\Pi_{\mu\nu}^P$ who belongs to $\hat T_1(t)$ of
the form
\begin{equation}\label{vm:eq:sandwich}
    [\bar u_1(ig\bm T_a )\gamma^\mu u_1]\frac{(-i)}{q^2}\
    i\Pi^P_{\mu\nu}\delta^{ab}\
    \frac{(-i)}{q^2} [\bar u_2(ig\bm T_b )\gamma^\nu u_2].
\end{equation}
\begin{center}\includegraphics[scale=0.5]{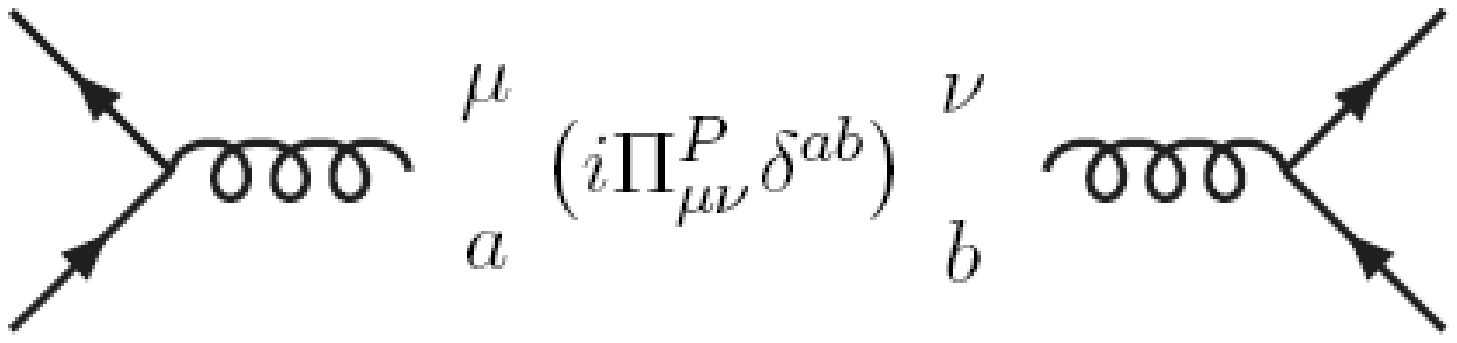}  \end{center}
The correct color factor is recover by using the antisymmetry of the structure constants
\begin{equation}\label{}
    2f^{abc}\bm T_b \bm T_c = f^{abc}
    \left[\bm T_b,\bm T_c\right]=
    if^{abc}f^{bcd}\bm T_d
    =iN\delta^{ad}\bm T_d.
\end{equation}
This relation is valid for any representation $\bm T^a$ of the Lie
algebra $\frak{su}(N)$ where $N$ is the color number. If we take the
adjoint representation $(\bm T^a)^{bc}=-if^{abc}$, we get another
relation
\begin{equation}\label{}
    2f^{axy}f^{bxz}f^{cyz} = Nf^{abc},
\end{equation}
useful when dealing with the 4-gluon vertex. To explicit calculate
the pinching contribution of the graph of Fig.~\ref{vm:fig:vertex},
it is convenient to decompose the vertices into two pieces. A piece
$\Gamma^F$ which has terms with external momentum $q$ and a piece
$\Gamma^P$ which carries the internal momentum only:
\begin{eqnarray}\label{vm:eq:vertex}
  \Gamma_{\alpha\mu\beta}(k,q)&=&
  (k-q)_\beta g_{\alpha\mu} + (k+2q)_\alpha g_{\mu\beta} - (2k+q)_\mu g_{\beta\alpha}\\\nonumber
  &=&\Gamma_{\alpha\mu\beta}^F(k,q) +  \Gamma_{\alpha\mu\beta}^P(k,q)\\\nonumber
  \Gamma_{\alpha\mu\beta}^F(k,q) &=& -(2k+q)_\mu g_{\alpha\beta} +2q_\alpha g_{\mu\beta} - 2q_\beta g_{\mu\alpha}\\\nonumber
  \Gamma_{\alpha\mu\beta}^P(k,q) &=& k_\alpha g_{\mu\beta} + (k+q)_\beta g_{\alpha\mu}
\end{eqnarray}
Now $\Gamma^{F}_{\mu\nu\alpha}$ satisfies a Feynman-gauge Ward
identity:
\begin{equation}
    q^{\mu}\Gamma^{F}_{\alpha\mu\beta} = [k^2-(k+q)^2]g_{\alpha\beta},
\end{equation}
where the Right Hand Side (RHS) is the difference of two inverse
propagators in the Feynman gauge. As for
$\Gamma^{P}_{\mu\nu\alpha}$, it gives rise to pinch parts when
contracted with $\gamma$ matrices
\begin{eqnarray}\label{vm:Pinch2}
 \nonumber
  g_{\mu\alpha}(q\!\!\!/+k\!\!\!/) &=&
  ig_{\mu\alpha}[S^{-1}(p+q)-S^{-1}(p-k)],\\
  g_{\nu\alpha}k\!\!\!/ &=&ig_{\nu\alpha}[S^{-1}(p)-S^{-1}(p-k)].
\end{eqnarray}
Both $S^{-1}(p+q)$ and $S^{-1}(p)$ vanish on shell, whereas the two
terms proportional to $S^{-1}(p-k)$ pinch out the internal fermion
propagator in graph of Fig.~\ref{vm:fig:vertex}, and so we are left
with two ``pinch" (propagator-like) parts and one ``regular" (purely
vertex-like) part, namely

\begin{eqnarray}
 \nonumber
  \includegraphics[width=3cm]{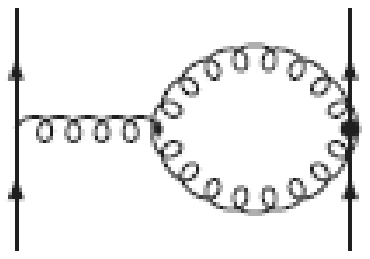} 
   &=&  i\Pi_{\mu\nu}^P \gamma^\mu\\
&=& Ng^2\int\frac{d^4k}{(2\pi)^4}\frac{1}{k^2(k+q)^2}(g_{\mu\nu}\gamma^{\nu}) \\
  \text{regular~part} &=& \frac{1}{2}Ng^2
    \int\frac{d^4k}{(2\pi)^4}\frac{\gamma^{\rho}S(p-k)
    \gamma^{\sigma}} {k^2(k+q)^2}{\Gamma}^{F}_{\rho\sigma\mu}
\end{eqnarray}

We can always add a $q_\mu$ term in the expression of the pinch part
since this term vanishes when contracted with the conserved current
of the quark $\bar u(p)\gamma^\mu u(p+q)$. The cancellation with the
self-energy is more evident if we write the total pinch contribution
of the vertex-like diagrams
\begin{equation}\label{vm:PinchVertex}
\boxed{i \Pi^P_{\mu\nu}(q)=2Ng^2 q^2t_{\mu\nu}(q)
\int\frac{d^4k}{(2\pi)^4}\frac{1}{k^2(k+q)^2}}
\end{equation}
A factor two comes from the mirror contribution and the $q^2$ is
added to recover the expression~\eqref{vm:eq:sandwich}. This
multiplicative factor $q^2$ will lead us in the following to the
rule of the intrinsic pinch technique.

\subsection{The intrinsic pinch technique}
\label{vm:sec:intrinsic}

Now, we have computed the pinch part coming from the graph (f) of
Fig.~\ref{vm:fig:qqscat}. We have to add this expression to the
conventional amplitude of \eqref{vm:SEoneloop}. This amplitude is
equal to (we omit the color factor $\delta^{ab}$)
\begin{equation}\label{}
        i\Pi_{\mu\nu}^{0} = \frac{Ng^2}{2(2\pi)^4}
        \int\frac{d^4k}{k^2(k+q)^2}
        \left[\Gamma_{\alpha\mu\beta}(k,q)\Gamma_{\beta\nu\alpha}(k+q,-q) -
        k_\mu(k+q)_\nu - k_\nu(k+q)_\mu\right],
\end{equation}
The gluon loop gives a symmetry factor $1/2$. The amplitude of ghost
loop diagram was symmetrized and the fermi statistique of the ghosts
leads to a minus sign. Note also that the expressions of the
vertices $\Gamma_{\alpha\mu\beta}(k,q)$ and
$\Gamma_{\beta\nu\alpha}(k+q,-q)$ are equal\footnote{These are the
same vertex with the opposite momenta but the minus sign is given by
the color factor.}. As previously, we decompose the vertices
$\Gamma_{\alpha\mu\beta}$ given by \eqref{vm:eq:vertex} into a
regular and a pinch parts. We rewrite the product of the two
vertices as\footnote{$\Gamma$ , $\Gamma^F$ and $\Gamma^P$ are
defined in \eqref{vm:eq:vertex}.}
\begin{equation}\label{vm:gammaFP}
    \Gamma_{\alpha\mu\beta}\Gamma_{\beta\nu\alpha} = \Gamma_{\alpha\mu\beta}^F\Gamma_{\beta\nu\alpha}^F +
    \Gamma_{\alpha\mu\beta}^P\Gamma_{\beta\nu\alpha} + \Gamma_{\alpha\mu\beta}\Gamma_{\beta\nu\alpha}^P
    -\Gamma_{\alpha\mu\beta}^P\Gamma_{\beta\nu\alpha}^P.
\end{equation}
Now, the momenta in $\Gamma^P$ trigger tree-level Ward identities on
the full vertex:
\begin{subequations}\label{vm:eq:WIvertex}
\begin{eqnarray}
  k^\alpha\Gamma_{\alpha\mu\beta}(k,q) &=& (k+q)^2t_{\mu\beta}(k+q) - q^2t_{\mu\beta}(q),\\
  (k+q)^\beta\Gamma_{\alpha\mu\beta}(k,q) &=& k^2t_{\mu\alpha}(k)- q^2t_{\mu\alpha}(q).
\end{eqnarray}
\end{subequations}
Where the transverse projector $t_{\mu\nu}$ was previously defined.
The first term of \eqref{vm:gammaFP} is saved in its entirety since,
as it generates no pinch. The fourth plays a role in canceling the
ghost loop
\begin{equation}\label{}
\Gamma_{\alpha\mu\beta}^P\Gamma_{\beta\nu\alpha}^P = k^2 g_{\mu\nu}
+ (k+q)^2g_{\mu\nu} + k_\mu(k+q)_\nu + k_\nu(k+q)_\mu,
\end{equation}
and the others two can be rewritten as
\begin{equation}\label{}
    \Gamma_{\alpha\mu\beta}^P\Gamma_{\beta\nu\alpha} + \Gamma_{\alpha\mu\beta}\Gamma_{\beta\nu\alpha}^P =
    -4q^2t_{\mu\nu}(q)+2k^2t_{\mu\nu}(k)+2(k+q)^2t_{\mu\nu}(k+q).
\end{equation}
We see that the first term is cancelled by the pinch contribution
\eqref{vm:PinchVertex}. In the ``intrinsic'' pinch technique,
introduce in Ref.~\cite{Cornwall:1989gv}, we simply drop this term
proportional to $q^2$. Indeed, the gauge dependence of the ordinary
graphs is cancelled by the contributions of the pinch graphs. Since
the pinch graphs are always missing one or more propagators
corresponding to the external legs, the gauge-dependent parts of the
ordinary graphs must also be missing one or more external propagator
legs. So if we extract systematically from the proper graphs the
part which are missing external propagator legs, i.e. proportional
to $q^2$, and simply throw them away, we obtain the gauge-invariant
results. Some others terms, like $k^2g_{\mu\nu}$, vanish by the
rules of the dimensional regularization \eqref{vm:DRrules} and the
gauge invariant gluon self-energy reads
\begin{equation}\label{vm:eq:GI_SE}
        i\widehat\Pi_{\mu\nu}(q) = \frac{Ng^2}{2(2\pi)^4}
        \int\frac{d^4k}{k^2(k+q)^2}
         \left[\Gamma_{\alpha\mu\beta}^F(k,q)\Gamma_{\beta\nu\alpha}^F(k+q,-q) -
        2(2k+q)_\mu(2k+q)_\nu\right].
\end{equation}
Note that this expression is the sum of a gluon-like and a
ghost-like contributions. Each contribution is separatively
transverse. We will see in the section \ref{vm:sec:BFM} that this
gauge-invariant gluon self-energy is exactly the self-energy of the
background gluon in BFM in Feynman gauge.

Of course, we want to go a step further and compute the integral.
To this end, we use the results
\begin{eqnarray*}
  \Gamma_{\alpha\mu\beta}^F\Gamma_{\beta\nu\alpha}^F &=& -8q^2t_{\mu\nu}(q)+4(2k+q)_\mu (2k+q)_\nu, \\
  \int\frac{d^4k}{k^2(k+q)^2} (2k+q)_\mu (2k+q)_\nu&=&
  \frac{1}{3}q^2t_{\mu\nu}(q)\int\frac{d^4k}{k^2(k+q)^2}.
\end{eqnarray*}
The second is easily obtained if we notice that the left hand side
is transverse, and hence proportional to $t_{\mu\nu}(q)$. The
coefficient is found by performing the trace. The gauge-independent
self-energy becomes
\begin{equation}\label{}
\boxed{
     i\widehat\Pi_{\mu\nu}^{ab}(q) =
     \frac{11}{3}Ng^2\delta^{ab}q^2t_{\mu\nu}(q)
        \int\frac{d^4k}{(2\pi)^4}\frac{1}{k^2(k+q)^2}}
\end{equation}
The integral is not convergent in 4 dimensions. The renormalization
scheme we use is the dimensional regularization since it preserves
the gauge-invariance of the theory. To regularize this integral we
need to introduce a Feynman parameter $l=k+xq$ and a arbitrary mass
$\mu$ to keep the interaction constant $g$ dimensionless. We find
\begin{eqnarray*}
   g^2\int\frac{d^4k}{(2\pi)^4}\frac{1}{k^2(k+q)^2}
   &=&\mu^{2\epsilon}g^2
   \int\frac{d^dl}{(2\pi)^d}\int_0^1\frac{dx}{(l^2-x(x-1)q^2)^2},\\
   &=&ig^2\int_0^1\frac{dx}{(4\pi)^{d/2}}\frac{\Gamma(\epsilon)}{\Gamma(2)}
    \left(\frac{\mu^2}{\Delta}\right)^{\epsilon},\\
    &=&i\frac{g^2}{(4\pi)^2}\int_0^1 dx
    \left(\frac{2}{\epsilon}+\ln 4\pi-\gamma+\ln\frac{\Delta}{\mu^2}
    +\cal O (\epsilon)\right),\\
    &=&i\frac{g^2}{16\pi^2}\ln\frac{q^2}{\mu^2},
\end{eqnarray*}
where we set $\Delta=x(x-1)q^2$. A factor $i$ arises when
performing a Wick rotation in the integral ($q^2>0$ since we are
working with Euclidean momenta). This $i$ factor cancel with the
one in the definition of $i\Pi_{\mu\nu}$. In our development we
followed the rules of the $\overline{MS}$. Finally the
gauge-invariant self-energy reads
\begin{equation}\label{vm:RunnCoupl}
    \widehat{\Pi}(q^2)= - bg^2q^2\ln(q^2/\mu^2),
\end{equation}
and $b= 11N/48\pi^2$, the coefficient in front of $(- g^3)$ in the
usual one loop $\beta$ function in a pure gauge theory. The
inclusion of fermions is straightforward. In a theory with $n_f$
quark flavours, the first coefficient of the $\beta$-function
becomes $b=(11N-2n_f)/48\pi^2$.

The final expression of the gauge-independent propagator is
\begin{equation}\label{vm:eq:GIpropagator}
    \boxed{\widehat\Delta^{-1}(q^2)=
    q^2\left[1+bg^2\ln\left(\frac{q^2}{\mu^2}\right)\right]}
\end{equation}
We see that the gluon remains massless in perturbation theory. The
dynamically generated mass of the gluon is a non-perturbative
feature of the non-abelian gauge theory
\cite{Cornwall:1982zr,Aguilar:2006gr}. Due to the abelian Ward
identities satisfied by the pinch technique effective Green's
functions, the renormalization constants of the gauge-coupling and
the effective self-energy satisfy the QED relation $Z_g=\widehat
Z_A^{1/2}$. Hence the product $\widehat
d(q^2)=g^2\widehat\Delta(q^2)$ forms a renormalization-group
invariant ($\mu$-independent) quantity for large momenta $q^2$,
\begin{equation}\label{}
    \widehat d(q^2) = \frac{\bar{g}^2(q^2)}{q^2}.
\end{equation}
$\bar g(q^2)$ is the renormalization-group invariant effective
charge of QCD,
\begin{equation}
\overline{g}^2(q^2) = \frac{g^2}{1+bg^2\ln(q^2/\mu^2)}
 = \frac{1}{b\ln(q^2/\Lambda_{QCD}^2)},
\end{equation}
with $\Lambda_{QCD}= \mu \exp[-1/(2bg^2)]$. The value,
$\Lambda_{QCD}\approx 300$ MeV, can be related to experimental
data and defines the limit of the validity of the perturbation
theory. It is worth mentioning that its value is actually scheme
and order dependent. This effective charge, defined as the
radiative corrections to the coupling constant, matches, for large
momenta, onto the running coupling constant, defined as the
solution of the renormalization group equation
\begin{equation}\label{}
    \mu\frac{\partial g}{\partial\mu}=\beta(g)=-bg^3,
\end{equation}
at the one-loop level.

\subsection{The gluon self-energy in a general covariant gauges}
The previous section showed how the pinch technique works. We took
the simplest example of the Feynman gauge, but the same technique
can be used in the general Lorentz gauges. In this case, momenta
are also present in the gluon propagator. And thus, the other
diagrams like (e) and (g) in Fig.~\ref{vm:fig:qqscat69} give a non
zero pinch contribution to the gluon self-energy. Of course, these
contributions are proportional to $\lambda=\xi-1$. The expressions
of these pinch parts can be found by simply using the tree-level
Ward identities \eqref{vm:eq:WIelem}.

\medskip

The pinch contribution of the box diagram (and its mirror graph) is
given by
\begin{equation}\label{vm:eq:Box}
B^P_{\mu\nu}(q) = \l
Nq^4\bigg[t_{\mu\nu}\int\frac{d^4k}{(2\pi)^4}\frac{1}{k^4(k+q)^2}
         +\frac{\l}{2}t_{\mu\rho}t_{\sigma\nu}\
        \int\frac{d^4k}{(2\pi)^4}\frac{k^\rho
        k^\sigma}{k^4(k+q)^4}\bigg].
\end{equation}
In this subsection, we did not write the $q$-dependence of the
transverse projector $t_{\mu\nu}\equiv t_{\mu\nu}(q)$. The graphs
such as (g) in Fig.~\ref{vm:fig:qqscat69} have a contribution,
\begin{equation}\label{}
    V_{1,\mu\nu}^P(q) = -\l
    Nq^2t_{\mu\nu}\int\frac{d^4k}{(2\pi)^4}\frac{1}{k^4},
\end{equation}
who vanishes by the rules of the dimensional regularization
\eqref{vm:DRrules}. We are left with the pinch part of the vertex
graph proportional to $\l$ :
\begin{equation}\label{}
    V_{2,\mu\nu}^P(q) = \l Nq^2\bigg[t_{\mu\nu}\int\frac{d^4k}{(2\pi)^4}\frac{k^2+4k\cdot q}{k^4(k+q)^2}
         -\l q^2t_{\mu\rho}t_{\sigma\nu}\
        \int\frac{d^4k}{(2\pi)^4}\frac{k^\rho
        k^\sigma}{k^4(k+q)^4}\bigg]. 
\end{equation}
The sum of all these terms are equal to the pinch part in the
general covariant gauges
\begin{eqnarray}
 \nonumber
  i\Pi^P_{\mu\nu}(q)|_{\xi\neq 1} &=& B^t_{\mu\nu}(q)+ V_{1,\mu\nu}^P(q) + V_{2,\mu\nu}^P(q),\\\nonumber
   &=& \l Nq^2\bigg[t_{\mu\nu}\int\frac{d^4k}{(2\pi)^4} \frac{2k\cdot
q}{k^4(k+q)^2} -\frac{\l}{2}q^2t_{\mu\rho}t_{\sigma\nu}\
        \int\frac{d^4k}{(2\pi)^4}\frac{k^\rho
        k^\sigma}{k^4(k+q)^4}\bigg], \\
   &=&-i\frac{\l}{4}(\l+8)Ng^2 t_{\mu\nu}.
\end{eqnarray}
You can indeed check that this expression is the opposite of the
gauge dependent part of the conventional self-energy.

\section{The Background Field Method}\label{vm:sec:BFM}
The background field method (BFM) is an elegant and powerful
formalism whereby gauge invariance of the generating functional is
preserved. The method was first introduced by DeWitt \cite{Dea},
and was extended by 't Hooft \cite{tH}, Boulware \cite{Bo} and
Abbot \cite{Abb}. In our exposition, we will follow the very
readable account of the last author.

In the first subsection, we present the generating functional for
the connected and irreducible Green's functions of the
conventional theory and in the background field method. The
equivalence of the background field method and the conventional
approach is developed in the second subsection. Finally, in the
last subsection, we recover the PT gauge-invariant self-energy by
the background field method.

\subsection{Path integral formalism}

Consider the generating functional for pure Yang-Mills field.
Fermions play no role in the background field method, they are
treated as in the ordinary formalism, and will be neglected. We
write it as
\begin{equation}\label{vm:Z}
    Z[J] = \int{\cal D}Q \det\left[\frac{\delta G^a}{\delta w^b}\right] \exp\left( i\int d^4x
    \left[{\cal L}(Q) -\frac{1}{2\xi}G_aG^a + J^\mu_a
    Q^a_\mu\right]\right),
\end{equation}
with the usual definitions :
\begin{eqnarray}\label{}
    {\cal L}(Q) &=& -\frac{1}{4} F_{\mu\nu}^a F^{\mu\nu}_a,\\
    F_{\mu\nu}^a&=& \partial_\mu Q_\nu^a-\partial_\nu Q_\mu^a + g
    f^{abc}Q_\mu^bQ_\nu^c.
\end{eqnarray}
$G^a$ is the gauge-fixing term, and in the covariant Lorentz gauges,
we have $G^a=\partial^\mu Q_\mu^a$. $\delta G^a/\delta w^b$ is the
derivative of the gauge-fixing term under an infinitesimal gauge
transformation
\begin{equation}\label{vm:eq:gaugetrans1}
    \delta Q^a_\mu = -f^{abc}\omega^b
    Q_\mu^c+\frac{1}{g}\partial_\mu\omega^a.
\end{equation}
Under this transformation, $F_{\mu\nu}^a$ becomes
$F_{\mu\nu}^a-f^{abc}\omega^bF_{\mu\nu}^c$ and thus ${\cal L}(Q)$ is
gauge-invariant. The functional derivatives of $Z[J]$ with respect
to $J$ are the disconnected Green functions of the theory. The
connected Green functions are generated by
    $W[J] = -i\ln Z[J]$.
Finally, one defines the effective action by making the Legendre
transformation
\begin{equation}\label{}
    \Gamma[\bar{Q}] = W[J]-\int d^4
    xJ^\mu_a\bar Q_\mu^a,\quad\text{where   }\quad
    \bar{Q}^a_\mu = \frac{\delta W}{\delta J^\mu_a}.
\end{equation}
The derivative of the effective action with respect to $\bar Q$ are
the one-particle-irreducible Green's functions of the theory.

We now define quantities analogous to $Z$, $W$, and $\Gamma$ in the
background field method. We denote these by $\tilde Z$, $\tilde W$,
and $\tilde \Gamma$\footnote{The quantities are written with a
$\tilde{ }\ $ in the background field and with a $\widehat{ }\ $ in
the pinch technique.}. They are define exactly like the conventional
generating functionals except that the field in the classical
lagrangian is written not $Q$ but as $A+Q$, where $A$ is the
background field. We do not couple the background field to the
source $J$. Thus, we define
\begin{equation}\label{vm:Ztilde}
\tilde Z[J,A] = \int{\cal D}Q \det\left[\frac{\delta \tilde
G^a}{\delta w^b}\right] \exp i\int d^4x \left[{\cal L}(A+Q)
-\frac{1}{2\xi_Q}\tilde G_a\tilde G^a + J^\mu_a
    Q^a_\mu\right],
\end{equation}
where $\delta\tilde G^a/\delta\omega^b$ is the derivative of the
gauge-fixing term under the infinitesimal gauge transformation
$\delta
Q^a_\mu=-f^{abc}\omega^b(A^c_\mu+Q^c_\mu)+(1/g)\partial_\mu\omega^a$.
As previously, $\xi_Q$ is an arbitrary parameter, and thus there are
a background Feynman gauge ($\xi_Q=1$), a background Landau gauge
($\xi_Q=0$), etc. Then, just as in the conventional approach, we
define
    $\tilde W[J,A]=-i\ln\tilde Z[J,A]$
and the background effective action
\begin{equation}\label{}
    \Gamma[\tilde{Q},A] = \tilde W[J,A]-\int d^4
    xJ^\mu_a\tilde Q_\mu^a,\quad\text{where   }\quad
    \tilde{Q}^a_\mu = \frac{\delta\tilde W}{\delta J^\mu_a}.
\end{equation}
Since there are several field variables being used here, it is
worthwhile to summarize them :
\begin{itemize}
  \item $Q_\mu^a=$ the quantum field, the variable of the
  integration in the functional formalism;
  \item $A_\mu^a=$ the background field;
  \item $\bar Q_\mu^a=\delta W/\delta J_\mu^a=$ the argument of the
  conventional effective action $\Gamma[\bar Q]$;
  \item $\tilde Q_\mu^a=\delta\tilde W/\delta J_\mu^a=$ the quantum field argument of
  the background field effective action $\tilde\Gamma[\tilde Q, A]$;
\end{itemize}
Since ${\cal L}(Q)$ is invariant under \eqref{vm:eq:gaugetrans1},
${\cal L}(A+Q)$ is invariant under
\begin{subequations}\label{vm:gaugetrans3}
\begin{eqnarray}\label{vm:gaugetrans2}
  \delta Q_\mu^a &=& -f^{abc}\omega^bQ_\mu^c, \\
  \delta A_\mu^a &=&
  -f^{abc}\omega^bA_\mu^c+\frac{1}{g}\partial_\mu\omega^a.
\end{eqnarray}
\end{subequations}
The transformation \eqref{vm:gaugetrans2} corresponds simply to a
change of variables in $\tilde{Z}[J,A]$. If we perform the
transformation
$\delta J_\mu^a = -f^{abc}\omega^bJ_\mu^c$,
and choose the background field gauge condition
\begin{equation}\label{vm:BFgauge}
    \tilde G^a=\partial^\mu Q_\mu^a+gf^{abc}A_\mu^bQ_\mu^c,
\end{equation}
such that $\tilde G_\mu^a\tilde G_a^\mu$ is invariant under
\eqref{vm:gaugetrans3}, we see that $\tilde Z[J,A]$ and $\tilde
W[J,A]$ are invariant. It then follows that $\tilde\Gamma[\tilde
Q,A]$ is invariant under the transformations
\begin{subequations}
\begin{eqnarray}
    \label{vm:delta_A1}
  \delta A_\mu^a &=& -f^{abc}\omega^bA_\mu+\frac{1}{g}\partial_\mu\omega^a, \\
  \delta\tilde Q_\mu^a &=& -f^{abc}\omega^b\tilde Q_\mu^c,
\end{eqnarray}
\end{subequations}
in the background field gauge. In particular, $\tilde\Gamma[0,A]$
must be an explicit gauge-invariant functional of $A$ since
\eqref{vm:delta_A1} is just an ordinary gauge transformation of the
background field. The quantity $\tilde\Gamma[0,A]$ is the
gauge-invariant effective action which one computes in the
background field method. In sect.~\ref{vm:sec:eqBFM}, it will be
shown that $\tilde\Gamma[0,A]$ is equal to the usual effective
action $\Gamma[\bar Q]$, with $\bar Q=A$, calculated in an
unconventional gauge which depends on $A$. Thus $\tilde\Gamma[0,A]$
can be used to generate the S-matrix of a gauge theory in exactly
the same way as the usual effective action is employed.


\subsection{Equivalence of the background field method}\label{vm:sec:eqBFM}
We now derive relationships between $Z$, $W$, $\Gamma$ and the
analogous quantities $\tilde Z$, $\tilde W$, and $\tilde\Gamma$ of
the background field method. This is done by making the change of
variables $Q\rightarrow Q-A$ in Eq. \eqref{vm:Ztilde}. One then
finds that when $\tilde Z[J,A]$ is calculated in the background
field gauge of Eq. \eqref{vm:BFgauge},
\begin{equation}\label{vm:Ztilde2}
    \tilde Z[J,A]=Z[J]\exp\left(-i\int d^4x J_a^\mu
    A_\mu^a\right),
\end{equation}
where $Z[J]$ is the conventional generation functional of
eq.~\eqref{vm:Z} evaluated with the gauge-fixing term
\begin{equation}\label{vm:BFgauge2}
    G^a=\partial^\mu Q_\mu^a-\partial^\mu
    A_\mu^a+gf^{abc}A_\mu^bQ_\mu^c.
\end{equation}
One can verify that the ghost determinant of $\tilde Z$ in the
background field gauge goes over into the correct ghost determinant
for $Z$ in the gauge of Eq. \eqref{vm:BFgauge2}. Note that because
of the presence of the background field $A$ in the gauge-fixing term
\eqref{vm:BFgauge2}, $\tilde W$ will be a functional of $A$ as well
as $J$. It follows from \eqref{vm:Ztilde2} that $W$ and $\tilde W$
are related by
\begin{equation}\label{vm:Wtilde}
    \tilde W[J,A]=W[J]-\int d^4xJ^\mu_aA_\mu^a.
\end{equation}
Like $Z[J]$, $W[J]$ depends on $A$ through the gauge-fixing term.
Taking a derivative of \eqref{vm:Wtilde} with respect to $J$ and
recalling that $\bar Q=\delta W/\delta J$ and $\tilde Q=\delta
\tilde W/\delta J$ we find that
\begin{equation}\label{}
    \tilde Q^a_\mu=\bar Q^a_\mu-A_\mu^a.
\end{equation}
Finally, performing a Legendre transformation on the relation
\eqref{vm:Wtilde} we have a relation between the background field
effective action and the conventional effective action
\begin{equation}\label{vm:Gammatilde}
    \tilde\Gamma[\tilde Q,A]=\Gamma[\bar Q]|_{\bar Q=\tilde Q+A} =
    \Gamma[\tilde Q+A].
\end{equation}
The gauge-invariant effective action is just $\tilde\Gamma[0,A]$ so
from \eqref{vm:Gammatilde} we  have the identity we need
\begin{equation}\label{vm:proofGamma}
    \tilde\Gamma[0,A]=\Gamma[\bar Q]|_{\bar Q=A}.
\end{equation}
In this identity, $\tilde\Gamma$ is calculated in the background
field gauge of eq. \eqref{vm:BFgauge} and $\Gamma$ in the gauge of
\eqref{vm:BFgauge2}. Thus, in eq. \eqref{vm:proofGamma}, $\Gamma$
depends on $A$ both through this gauge-fixing term and because $\bar
Q=A$.

The gauge-invariant effective action, $\tilde{\Gamma}[0,A]$, is
computed by summing all one-particle irreducible diagrams with $A$
fields on external legs and $Q$ field inside loops. No $Q$ field
propagators appear on external lines (since $\tilde Q=0$) and
likewise no $A$ field propagators occur inside loops (since the
functional integral is only over $Q$). Note that because $A$ appears
in the gauge condition \eqref{vm:BFgauge2} and acts as a source
there, the one-particle-irreducible Green's functions calculated
from the gauge-invariant effective action will be very different
from those calculated by the conventional methods in normal gauges.
Nevertheless, the relation \eqref{vm:proofGamma} assures us that all
gauge-independent physical quantities will come out the same in
either approach. Because the effective action involves only
one-particle-irreducible diagrams, vertices with only line outgoing
quantum line will never contribute. The Feynman rules in the
background field method are given in appendix B.

\subsection{The self-energy of the background gluon}
\begin{figure}
\begin{center}
\includegraphics[width=10cm]{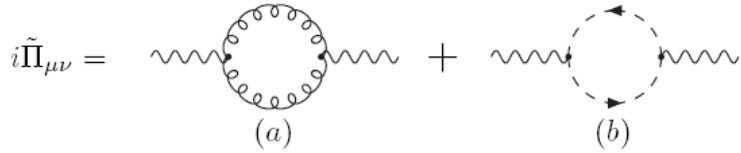}
\caption{\label{vm:fig:BFMgluonSE}Background gluon self energy.}
\end{center}
\end{figure}

With the Feynman rules given in appendix B, we can compute easily
the expression of the gluon self-energy. This is the sum of a gluon
and a ghost contribution. We only display these two relevant graphs
on Fig.~\ref{vm:fig:BFMgluonSE} since the two others vanish by the
rules of the dimensional regularization. Before evaluating the
amplitude of the remaining diagrams, let us remark that the
three-point vertex with one background field, define as
$\tilde{\Gamma}^{abc}_{\alpha\mu\beta}=
gf^{abc}\tilde{\Gamma}_{\alpha\mu\beta}$ with
\begin{equation}\label{}
\tilde{\Gamma}_{\alpha\mu\beta}(p,q,r) = \left(p-q+ \frac{1}{\xi_Q}r
\right)_\beta g_{\alpha\mu} + \left(q-r-\frac{1}{\xi_Q}p
\right)_\alpha g_{\mu\beta} + (r-p)_\mu g_{\alpha\beta},
\end{equation}
correspond to the expression of the $\Gamma_{\alpha\mu\beta}^F$ in
eq.~\eqref{vm:eq:vertex} when it is evaluated in the background
Feynman gauge $\xi_Q=1$, i.e.
\begin{equation}
\tilde{\Gamma}_{\alpha\mu\beta}(k,q,-q-k)|_{\xi_Q=1}=-(2k+q)_\mu
g_{\alpha\beta} +2q_\alpha g_{\mu\beta} - 2q_\beta g_{\mu\alpha}.
\end{equation}
This fact gives a hint that the background field method may
reproduce the same results which are obtained by the pinch
technique.

Now, we calculate the gluon self-energy in the background field
method with the background Feynman gauge. Amplitudes of the diagrams
\ref{vm:fig:BFMgluonSE}(a) and \ref{vm:fig:BFMgluonSE}(b) give the
contributions
\begin{subequations}
\begin{eqnarray}\label{}
i\widehat{\Pi}^{(a)}_{\mu\nu} (q) &=& \frac{Ng^2}{2}
        \int\frac{d^4k}{(2\pi)^4}\frac{1}{k^2(k+q)^2}
         \Gamma_{\alpha\mu\beta}^F(k,q)\Gamma_{\beta\nu\alpha}^F(k+q,-q),\\
i\widehat\Pi_{\mu\nu}^{(b)} &=& -Ng^2
        \int\frac{d^4k}{(2\pi)^4}\frac{1}{k^2(k+q)^2}
         (2k+q)_\mu(2k+q)_\nu,
\end{eqnarray}
\end{subequations}
which correspond respectively to the first and second terms in the
expression of the gauge-invariant self-energy
$\widehat{\Pi}_{\mu\nu}$ of Eq. \eqref{vm:eq:GI_SE}. We proved, by a
explicit calculations, that the pinch technique gauge-invariant
self-energy of the gluon can be recover easier by the background
field method in the Feynman gauge at the one-loop level. We now
continue our analysis and show that this equivalence still holds for
the 3-gluon and the 4-gluon vertex.

\section{Gauge-invariant gluon vertex}
\label{vm:sec:3and4gluons}
\subsection{Gauge-independent three-gluon
vertex with the intrinsic pinch technique}\label{vm:sec:3gluons}
The calculation of the gauge-invariant three-gluon vertex by the
S-matrix pinch technique is much more tedious that for the
propagator. The road map of the way the vertex is constructed is
given in \cite{Cornwall:1989gv}. But here, we shall just explain the
construction of the vertex by the intrinsic pinch technique.
\begin{figure}
\begin{center}
\includegraphics[width=10cm]{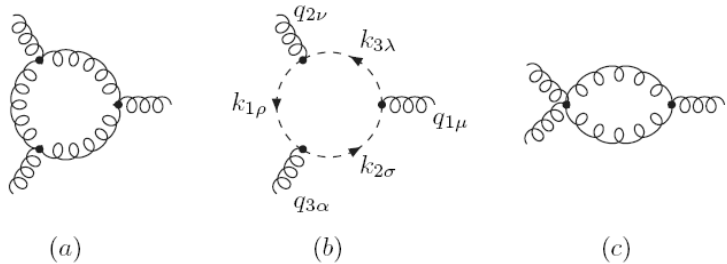}
\caption{\label{vm:fig:Vertex3g} One loop graphs for the conventional 3-gluon vertex.}
\end{center}
\end{figure}
The relevant graphs for the three-gluon vertex at the one-loop level
are depicted in Fig.~\ref{vm:fig:Vertex3g}. The contribution of the
gluon loop and the symmetrized ghost loop read
\begin{eqnarray}\label{}
\Gamma^{\ref{vm:fig:Vertex3g}(a)}_{\mu \nu \alpha} &=&
    \frac{iNg^{2}}{2(2\pi)^4} \int \frac{d^4k}{k^2_1 k^2_2 k^2_3}
    \Gamma_{\sigma \mu \lambda}(k_2, q_1, -k_3)
    \Gamma_{\lambda \nu \rho}(k_3, q_2, -k_1)
    \Gamma_{\rho \alpha \sigma}(k_1, q_3, -k_2),\\
\Gamma^{\ref{vm:fig:Vertex3g}(b)}_{\mu \nu \alpha} &=&
    -\frac{iNg^{2}}{2(2\pi)^4} \int \frac{d^4k}{k^2_1 k^2_2 k^2_3}
    \left(k_{1\nu}k_{2\alpha}k_{3\mu}+k_{1\alpha}k_{2\mu}k_{3\nu}\right),
\end{eqnarray}
where we omit the group theoretical factor $gf^{abc}$. The momenta
and Lorentz indices are defined in Fig.~\ref{vm:fig:Vertex3g}(b).
All momenta $q_i$ are incoming such that $k_3=k$, $k_1=k+q_2$, and
$k_2=k-q_1$. As previously, we rewrite the 3-gluon vertices in
$\Gamma^P+\Gamma^F$ form. We let the $\Gamma^P$ generate Ward
identities and drop the terms proportional to $q^2_i$ . The
numerator of the gluon loop amplitude can be written as
\begin{equation}\label{vm:eq:3g_gamma}
  \Gamma^F_1\Gamma^F_2\Gamma_3^F+ \Gamma^P_1\Gamma_2\Gamma_3
  +  \Gamma_1\Gamma^P_2\Gamma_3 +  \Gamma_1\Gamma_2\Gamma_3^P -
  \Gamma^P_1\Gamma^P_2\Gamma_3  
  - \Gamma^P_1\Gamma_2\Gamma_3^P -
   \Gamma_1\Gamma^P_2\Gamma_3^P +  \Gamma^P_1\Gamma^P_2\Gamma_3^P.
\end{equation}
Here each vertex labeled $1$ carried the indices $\sigma\mu\lambda$,
each vertex labeled $2$ carried the indices $\lambda\nu\rho$, and
each vertex labeled $3$ carried the indices $\rho\alpha\sigma$ ;
1,2,3 refer to the exterior momentum labels. For instance, the first
term on the RHS of \eqref{vm:eq:3g_gamma} really means
\begin{equation}\label{}
\Gamma^F_{\sigma\mu\lambda}(k_2,q_1,-k_3)
\Gamma^F_{\lambda\nu\rho}(k_3,q_2,-k_1)
\Gamma^F_{\rho\alpha\sigma}(k_1,q_3,-k_2).
\end{equation}
As with the propagator, the first term on the RHS of
\eqref{vm:eq:3g_gamma} contains no pinch. Each of the next six terms
has pinches (i.e. term in $q^2_i$) coming from the action of
$\Gamma^P$ on the full vertex $\Gamma$, via the Ward identities
\eqref{vm:eq:WIvertex}. Some terms can refer to an internal momentum
$k_i^2$, in which case they give rise to an integral with only two
propagators. Let us note that the last term in
\eqref{vm:eq:3g_gamma}, with three $\Gamma^P$'s,
\begin{equation}\label{}
    \begin{split}
(\Gamma^P_1\Gamma^P_2\Gamma_3^P)_{\mu\nu\alpha}= &
    k_3^2(k_{2\mu}g_{\alpha\nu}+k{_{1\nu}}g_{\alpha\mu})
     + k_2^2(k_{3\mu}g_{\alpha\nu}+k{_{1\nu}}g_{\alpha\mu})\\
     &+ k_1^2(k_{2\mu}g_{\alpha\nu}+k{_{3\nu}}g_{\alpha\mu})
         +k_{1\nu}k_{2\alpha}k_{3\mu}+k_{1\alpha}k_{2\mu}k_{3\nu},
    \end{split}
\end{equation}
yields terms who cancel exactly the ghost contribution since the
three first terms vanish by symmetric integration. When we drop the
$q_i^2$ generated by the Ward identity we find the expressions for
the others of Eq. \eqref{vm:eq:3g_gamma}. We have
\begin{eqnarray}\nonumber
(\Gamma^P_1\Gamma_2\Gamma_3)_{\mu\nu\alpha} &=&
    k_1^2\left[\Gamma_{\nu\alpha\mu}(k_1,q_3,-k_2)+
    \Gamma_{\mu\nu\alpha}(k_3,q_1,-k_1)\right]\\
    &&-k_2^2k_{1\nu}g_{\alpha\nu}-k_3^2k_{1\alpha}g_{\mu\nu} -
    k_{1\nu}k_{2\alpha}k_{2\mu}+k_{1\alpha}k_{3\mu}k_{3\nu},\\
-(\Gamma^P_1\Gamma_2^P\Gamma_3)_{\mu\nu\alpha}&=&
    -k_3^2\Gamma_{\nu\alpha\mu}(k_1,q_3,-k_2) +
    k_2^2t_{\alpha\mu}(k_2)k_{3\nu} +
    k_1^2t_{\alpha\nu}(k_1)k_{3\mu}.
\end{eqnarray}
Finally, it remains to add the contribution of the
graph~\ref{vm:fig:Vertex3g}(c) and the two similar ones (by legs
permutation). The expression of the gauge-invariant 3-gluon vertex
at the one-loop level is then the sum of three terms
\begin{equation}\label{vm:eq:vert3gPT}
\widehat{\Gamma}_{\mu \nu \alpha}(q_1, q_2, q_3) =
    \Gamma_{\mu\alpha\nu}+
       \frac{iNg^{2}}{2(2\pi)^4}\int
       \frac{d^4 k}{k^2_1\ k^2_2\ k^2_3}\left(A_{\alpha\mu\nu} +
       B_{\alpha\mu\nu}\right)
       - C_{\alpha\mu\nu}.
\end{equation}
Introducing the notation
\begin{equation}\label{vm:Ai}
A(i)=\frac{iNg^{2}}{2(2\pi)^4}\int\frac{d^4 k}{k^2(k+q_i)^2},
\end{equation}
we will see in the following that each of the three terms
\begin{eqnarray}
 A_{\alpha\mu\nu}&=&\Gamma^F_{\sigma \mu \lambda}(k_2,
q_1) \Gamma^F_{\lambda \nu \rho}(k_3, q_2)
\Gamma^F_{\rho \alpha \sigma}(k_1, q_3), \nonumber \\
B_{\alpha\mu\nu} &=&2 (k_2+k_3)_{\mu}(k_3+k_1)_{\nu}(k_1+k_2)_{\alpha},\nonumber \\
C_{\alpha\mu\nu} &=&8(q_{1\alpha}g_{\mu \nu}- q_{1\nu}g_{\mu\alpha})
    A(1)+8(q_{2\mu}g_{\alpha\nu}-q_{2\alpha}g_{\mu\nu})A(2) \nonumber\\
    &&+8 (q_{3\nu}g_{\mu \alpha}-q_{3\mu}g_{\nu \alpha}) A(3) \nonumber,
\end{eqnarray}
is equal to a amplitude of a graph in the background field method.
We presented here only the ghost and gluon contributions to the
one-loop amplitude but the inclusion of fermion and scalar loops
are straightforward. Binger and Brodsky found that the forms
factor in $d$ dimensions of each contributions satisfy a relation
very closely linked to supersymmetry \cite{Binger:2006sj}.

\subsection{Gauge-independent three-gluon
vertex with the background field method}

\begin{figure}
\begin{center}
\includegraphics[width=14cm]{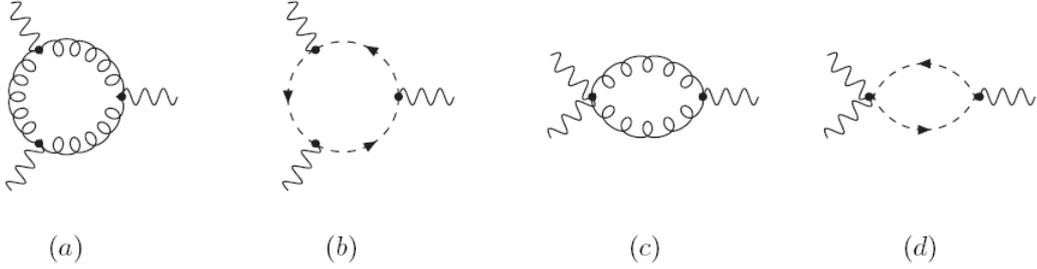}
\caption{\label{vm:fig:Vertex3gBFM} One loop graph for the 3-gluon vertex in the background field
method.}
\end{center}
\end{figure}

We saw in the previous section that the one-loop gluon self-energy
derived with the pinch technique can also be obtained, in a easier
way, by the background field method in the Feynman gauge. We, now,
shall show that the background field method can also be applied to
obtain the PT gauge-invariant three gluon vertex at the one-loop
level.

The relevant diagrams are shown in Fig.~\ref{vm:fig:Vertex3gBFM},
where the conventions for momenta and Lorentz and color indices are
displayed in Fig.~\ref{vm:fig:Vertex3g}. With the fact that an $AQQ$
vertex in the Feynman gauge, $\widetilde{\Gamma}_{\xi_Q=1}$, is
equivalent to $\Gamma^F$, it is easy to show that the contribution
of the diagram~\ref{vm:fig:Vertex3gBFM}(a) is
\begin{equation}\label{}
   \Gamma^{5(a)}_{\mu \nu \alpha}(q_1, q_2, q_3) =
     \frac{iNg^{2}}{2(2\pi)^4} \int
           \frac{d^4 k}{k^2_1 k^2_2 k^2_3}
      \Gamma^F_{\sigma \mu \lambda}(k_2, q_1)
         \Gamma^F_{\lambda \nu \rho}(k_3, q_2)
                \Gamma^F_{\rho \alpha \sigma}(k_1, q_3).
\end{equation}
The contribution of the diagram~\ref{vm:fig:Vertex3gBFM}(b) (and the
similar one with the ghost running the other way) is
\begin{equation}\label{}
 \Gamma^{5(b)}_{\mu \nu \alpha}(q_1, q_2, q_3) =
    \frac{iNg^{2}}{(2\pi)^4} \int
           \frac{d^4 k}{k^2_1 k^2_2 k^2_3}
    (k_2+k_3)_{\mu}(k_3+k_1)_{\nu}
         (k_1+k_2)_{\alpha} .
\end{equation}
When we calculate the diagram \ref{vm:fig:Vertex3gBFM}(c), again we
use the Feynman gauge ($\xi_Q=1$) for the four-point vertex with two
background fields. Remembering that the
diagram~\ref{vm:fig:Vertex3gBFM}(c) has a symmetric factor
$\frac{1}{2}$ and adding the two other similar diagrams, we find
\begin{equation}\label{}
\begin{split}
\Gamma^{5(c)}_{\mu \nu \alpha}(q_1, q_2, q_3) =  &
 8 (q_{1\alpha}g_{\mu \nu}- q_{1\nu}g_{\mu \alpha})
     A(1)+8 (q_{2\mu}g_{\alpha \nu}- q_{2\alpha}g_{\mu \nu})A(2)\\
    &  +8 (q_{3\nu}g_{\mu \alpha}- q_{3\mu}g_{\nu \alpha})A(3).
\end{split}
\end{equation}

Finally, the contribution of the diagram \ref{vm:fig:Vertex3gBFM}(d)
(and two other similar diagrams) turns out to be null because of the
group-theoretical identity for the structure constants $f^{abc}$
such as
\begin{equation}\label{}
    f^{ead} ( f^{dbx} f^{xce}+f^{dcx}f^{xbe}) = 0.
\end{equation}

Now adding the contributions from the diagrams (a)-(c) in
Fig.~\ref{vm:fig:Vertex3gBFM} and omitting the overall
group-theoretic factor $gf^{abc}$, we find that the result coincides
with the expression of Eq.(\ref{vm:eq:vert3gPT}) which was obtained
by the intrinsic pinch technique. Also we note that each
contribution from the diagrams (a)-(c), respectively, corresponds to
a particular term in Eq.(\ref{vm:eq:vert3gPT}).

We close this section with a mention that the constructed
$\widehat{\Gamma}_{\mu \nu \alpha}(q_1, q_2, q_3)$ is related to the
gauge-invariant propagator $\widehat{\Delta}$ of
Eq.\eqref{vm:eq:GIpropagator} through a Ward identity
\begin{equation}\label{vm:eq:WIvertex3g}
   q_1^{\mu}\widehat{\Gamma}_{\mu \nu \alpha}(q_1, q_2, q_3)
   =t_{\nu\alpha}(q_2)\widehat{\Delta}^{-1}(q_2)-
   t_{\nu\alpha}(q_3)\widehat{\Delta}^{-1}(q_3),
\end{equation}
which is indeed a naive extension of the tree-level one. It is very
important to note that the Ward identity makes no reference to ghost
Green's functions as the usual covariant-gauge Ward identities do.
Finally, we note that the RHS of \eqref{vm:eq:WIvertex3g} is not a
difference of two inverse propagators, because the projection
operators $t_{\mu\nu}$ has no inverse.

\subsection{Gauge-invariant four-gluon vertex}\label{vm:sec:4gluons}
The construction of the gauge-invariant 4-gluon vertex at the
one-loop by the S-matrix pinch technique is described in
\cite{Papavassiliou:1992ia}. This is a tedious work because of the
number of the graphs. In addition, new complications arise from
the fact that one-particle-reducible and one-particle-irreducible
graphs exchange contributions in a nontrivial way. We do not
report here the (very) lengthy expression of the vertex but, in
his paper, Papavassiliou shows that it obeys the Ward identity
\begin{equation}\label{vm:eq:PT4gluon}
\begin{split}
  q_1^\mu\widehat{\Gamma}_{\mu\nu\alpha\beta}^{abcd}(q_1,q_2,q_3,q_4)
  =&
  f^{abx}\widehat{\Gamma}^{cdx}_{\nu\alpha\beta}(q_1+q_2,q_3,q_4) \\
  &+ f^{acx}\widehat{\Gamma}^{bdx}_{\nu\alpha\beta}(q_2,q_1+q_3,q_4)\\
 &+ f^{adx}\widehat{\Gamma}^{bcx}_{\nu\alpha\beta}(q_2,q_3,q_1+q_4).
\end{split}
\end{equation}
which is a extension of the tree-level one. In this equation, in the
same way we defined the 3-gluon vertex
$g\widehat{\Gamma}^{cdx}_{\nu\alpha\beta}$, we defined the 4-gluon
vertex by $-ig^2\widehat{\Gamma}_{\mu\nu\alpha\beta}^{abcd}$.

The construction of the gauge-invariant 4-gluon vertex was also done
in \cite{Hashimoto:1994ct} in the context of the background field
method. The relevant graph and their amplitude can be found in this
reference. It is also proved that the 4-gluon vertex satisfies the
same Ward identity that the pinch technique one. Hence we are lead
to the conclusion that (at least) the longitudinal parts of the two
vertex $\tilde{\Gamma}_{\mu\nu\alpha\beta}^{abcd}$ (with 4
background gluons) and $\widehat{\Gamma}_{\mu\nu\alpha\beta}^{abcd}$
are equal.

\section{First principles and mathematical tools}
\label{sec:firstprinciples}

 Quantum field theory is based on fundamental principles,
such as the conservation of probability, causality, analyticity or
gauge invariance. Using these assumptions, we shall derive
constraints on the Green's functions of the theory, namely the
dispersion relations, the optical theorem and the Ward identities.

\subsection{Analyticity and renormalization}

Analyticity is one of the most important properties that governs
physical transition amplitudes. Correlation functions are considered
to be analytic in their kinematic variables, which is expressed by
means of the so-called Dispersion Relations (DRs)~\cite{KK,DR,ELOP}.
They were first derived in optics as a consequence of analyticity
and causality. In this section, we briefly review some important
facts about DRs and renormalization and discuss the subtleties
encountered in non-Abelian gauge theories.

\begin{figure}[htb]
\begin{center} 
\includegraphics[width=8cm]{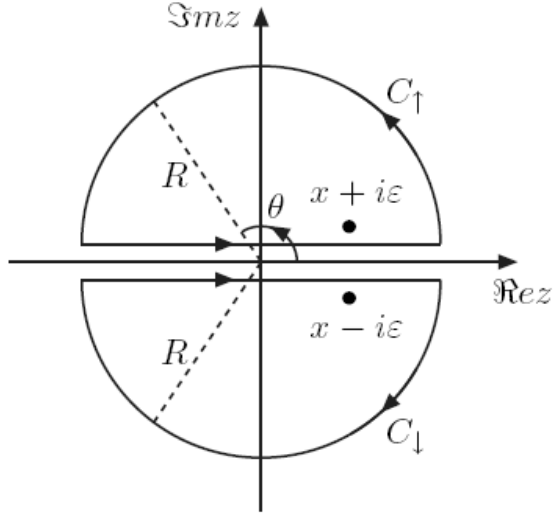}
\end{center}
\caption{\label{vm:fig:contour}Contours of complex integration.}
\end{figure}

If a complex function $f(z)$ is analytic in the interior of and upon
a closed curve, $C_\uparrow$ shown in Fig.~\ref{vm:fig:contour}, and
$x+i\varepsilon$ (with $x,\varepsilon \in$ {\bf R} and $\varepsilon
>0$) is a point within the closed curve $C_\uparrow$, we then have
the Cauchy's integral form,
\begin{equation}
f(x+i\varepsilon)\ =\ \frac{1}{2\pi i} \oint_{C_\uparrow} dz\,
\frac{f(z)}{z-x-i\varepsilon}\ ,
\end{equation}
where $\oint$ denotes that the path $C_\uparrow$ is singly wound.
Using Schwartz's reflection principle, one also obtains
\begin{equation}
f(x-i\varepsilon)\ =\ -\, \frac{1}{2\pi i} \oint_{C_\downarrow}
dz\, \frac{f(z)}{z-x+i\varepsilon}\ .
\end{equation}
Note that $C_\uparrow^*=C_\downarrow$. Sometimes, an analytic
function is called holomorphic; both terms are equivalent for
complex functions.

Of significant importance in the discussion of physical processes is
a DR, which relates the imaginary part of an analytic function
$f(x)$ to its real part, and vice versa. We assume for the moment
that the analytic function $f(z)$ has the asymptotic behaviour
$|f(z)|\le C/R^k$, for large radii $R$, where $C$ is a real
nonnegative constant and $k>0$; this assumption will be relaxed
later on, giving rise to more involved DR.  Taking now the limit
$\varepsilon\to 0$, it is easy to evaluate $\Re e f(x)$ through
\begin{equation}
2\Re e f(x)\ =\ `\lim_{\varepsilon\to 0}\mbox{'}\Big[
f(x+i\varepsilon) + f^*(x-i\varepsilon)\Big]\ =\
`\lim_{\varepsilon\to 0}\mbox{'}\,
\frac{1}{\pi}\int\limits_{-\infty}^{+\infty}dx'\, \Im m \left(
\frac{f(x')}{x'-x-i\varepsilon}\right)\, +\, \Gamma_{\infty}.
\end{equation}
Here, $`\lim_{\varepsilon\to 0}\mbox{'}$ means that the limit should
be taken \bleu{after} the integration has been performed, and
\begin{equation}
\Gamma_\infty\ =\ \frac{1}{\pi}\lim_{R\to \infty}\,
\Re e\, \int_0^\pi d\theta\, f(Re^{i\theta})\, .
\end{equation}
Because of the assumed asymptotic behaviour of $f(z)$ at infinity,
the integral over the upper infinite semicircle in Fig.\ 1,
$\Gamma_{\infty}$, can be easily shown to vanish. Employing the
well-known identity for distributions (the symbol P in front of the
integral stands for principle value integration),
\begin{displaymath}
`\lim_{\varepsilon\to 0}\mbox{'}\, \frac{1}{x'-x-i\varepsilon}\ =\
\mbox{P}\frac{1}{x'-x}\ +\ i\pi\delta (x'-x),
\end{displaymath}
we arrive at the unsubtracted dispersion relation,
\begin{equation}
\label{vm:DR1} \Re e f(x)\ =\ \frac{1}{\pi}\, \mbox{P}\,
\int\limits^{+\infty}_{-\infty} dx'\, \frac{\Im m f(x')}{x'-x}\ .
\end{equation}
Following a similar line of arguments, one can express the imaginary
part of $f(x)$ as an integral over $\Re e f(x)$.

In the previous derivation, the assumption that $|f(z)|$ approaches
zero sufficiently fast at infinity has been crucial, since it
guarantees that $\Gamma_{\infty}\to 0$. However, if we were to relax
this assumption, additional subtractions need to be included in
order to arrive at a finite expression. For instance, for $|f(z)|\le
CR^k$ with $k<1$, it is sufficient to carry out a single subtraction
at a point $x=a$. In this way, one has
\begin{equation}
\label{vm:DR2} \Re e f(x)\ =\ \Re e f(a)\ +\ \frac{(x-a)}{\pi}\,
\mbox{P}\, \int\limits_{-\infty}^{+\infty}dx'\, \frac{\Im m
f(x')}{(x'-a)(x'-x)}\ .
\end{equation}
From Eq.~\eqref{vm:DR2}, $\Re e f(x)$ can be obtained from $\Im m
f(x)$, up to a unknown, real constant $\Re e f(a)$. Usually, the
point $a$ is chosen in a way such that $\Re e f(a)$ takes a specific
value on account of some physical requirement.  For example, if $\Im
m f(q^2)$ is the imaginary part of the magnetic form factor of an
electron with photon virtuality $q^2$, one can prescribe that the
physical condition $\Re e f(0)=0$ should hold true in the Thomson
limit.

We next focus on the study of some crucial analytic properties of
off-shell transition amplitudes within the context of renormalizable
field theories. In such theories, one is allowed to have at most two
subtractions for a two-point correlation function. If $\Pi (s=q^2)$
is the self-energy function of a scalar particle with mass $m$ and
off-shell momentum $q$ ---the fermionic or vector case is
analogous---  then the real (or dispersive) part of this amplitude
can be fully determined by its imaginary (or absorptive) part via
the expression
\begin{equation}
\label{vm:DR3} \Re e \Pi (s)\ =\ \Re e \Pi (m^2)\, +\, (s-m^2)\Re e
\Pi'(m^2)\, +\, \frac{(s-m^2)^2}{\pi}\mbox{P}\,
\int\limits_{0}^{+\infty} ds'\, \frac{\Im m \Pi
(s')}{(s'-m^2)^2(s'-s)}\ .\quad
\end{equation}
${}$From Eq.\ (\ref{vm:DR3}), one can readily see that the two
subtractions, $\Re e \Pi (m^2)$ and the derivative $\Re e
\Pi'(m^2)$, respectively correspond to the mass and wave-function
renormalization constants in the on-mass shell scheme. At higher
orders, internal renormalizations of $\Im m \Pi (s)$, due to
counterterms coming from lower orders, should also be taken into
account. Then, Eq.\ (\ref{vm:DR3}) is still valid, {\em i.e.}, it
holds to order $n$ provided $\Im m \Pi (s)$ is renormalized to
order $n-1$. In general, the function $\Im m \Pi (s)$ has its
support in the non-negative real axis, {\em i.e.}, for $s\ge 0$.
This can be attributed to the semi-boundness of the spectrum of
the Hamiltonian, Spec$H\ge 0$~\cite{Khalfin}. Note that for
spectrally represented two-point correlation functions, we have
the additional condition $\Im m \Pi( m^2 )\ge 0
~$\cite{Bjorken,Justin}.

As has been mentioned above, in renormalizable field theories it
is required that $\Pi (s)$ should be finite after two subtractions
have been performed. This implies that
\begin{equation}
\label{vm:asympt} |\Pi (s) |\ \le\  C s^k\, ,\qquad \mbox{with}\quad
k<2,
\end{equation}
as $s\to \infty$. Obviously, the same inequality holds true for the
real as well as the imaginary part of $\Pi (s)$. In pure non-abelian
Yang-Mills theories, such as quark-less QCD, the transverse part,
$\Pi_T(s)$, of the gluon vacuum polarization behaves asymptotically
as, see Eq. \eqref{vm:RunnCoupl},
\begin{displaymath}
\Pi_T (s)\ \to\ C\, s\Big(\ln \frac{s}{\mu^2}\Big)^n\ .
\end{displaymath}
This result is consistent with Eq.\ (\ref{vm:asympt}), for any $n <
\infty$. Furthermore, we mention in passing that the
Froissart--Martin bound \cite{Froissart:ux},
\begin{equation}
\label{vm:Frois} |\Pi (s)|\ \leq\ C\,
s^3\Big(\ln\frac{s}{s_0}\Big)^2,
\end{equation}
at $s\to \infty$, which may be derived from axiomatic methods of
field theory, is weaker than Eq.\ (\ref{vm:asympt}). In fact, the
Froissart-Martin bound \cite{Froissart:ux} refers to the
asymptotic behaviour of a total cross section, $\sigma (s)$, in
the limit $s\to \infty$. This is expressed as $\sigma (s)\, \le\,
C [\ln(s/s_0)]^2$. Furthermore, the optical theorem gives the
relation $s\, \sigma(s) = \Im m T(s)$, where $T(s)$ is the
forward-scattering amplitude. If one assumes the absence of
accidental cancellations between the two-point function, $\Pi
(s)$, and higher $n$-point functions within the expression $\Im m
T(s)$, one can derive that
\begin{equation*}
|\Im m\Pi (s)|\, \leq\, C s^2 \Im m T(s)\, \leq\,  Cs^3
[\ln(s/s_0)]^2.
\end{equation*}
Because of analyticity, the $s$-dependence of $\Im m \Pi(s)$ will
affect the high-$s$ behaviour of $|\Pi(s)|$. Even if we assume that
the $s$-dependence thus induced on $|\Pi(s)|$ is the most modest
possible, {\em i.e.}, $|\Pi(s)| \sim \Im m \Pi(s)$ as $s
\rightarrow\infty $, still the tightest upper bound one could obtain
from these considerations is that of Eq.\ (\ref{vm:Frois}). The
analytic expression of gluon vacuum polarization satisfies Eq.\
(\ref{vm:Frois}). As a counter-example to this situation, we may
consider the Higgs self-energy in the unitary gauge; the absorptive
part of the Higgs self-energy has an $s^2$ dependence at high
energies, and its resummation~\cite{VW} is therefore not justified.

In the context of gauge field theories, one should anticipate a
similar analytic structure for two-point correlation functions.
However, an extra complication appears in such theories when
off-shell transition amplitudes are considered. In a theory with
spontaneous symmetry breaking, such as the Standard Model for
example, this complication originates from the fact that, in
addition to the physical particles of the spectrum of the
Hamiltonian, unphysical gauge dependent degrees of freedom, such as
would-be Goldstone bosons and ghost fields make their appearance.
Although on-shell transition amplitudes contain only the physical
degrees of freedom of the particles involved on account of
unitarity, their continuation to the off-shell region is ambiguous,
because of the presence of unphysical Landau poles, introduced by
the aforementioned unphysical particles. A reasonable prescription
for accomplishing such an off-shell continuation, which is very
close in spirit to the previous example of the scalar theory, would
be to continue analytically an off-shell amplitude by taking only
\bleu{ physical} Landau singularities into account.

Consider for example the off-shell propagator of a gauge particle in
the conventional $R_\xi$~gauges or BFGs, which runs inside a quantum
loop,
\begin{equation}
\label{vm:oprop1} \Delta_{0\mu\nu}^{(\xi_Q)}(q)\ =\ t_{\mu\nu}(q)\,
\frac{-i}{q^2-M^2}\, -\, \ell_{\mu\nu}(q)\, \frac{i\xi_Q}{q^2-\xi_Q
M^2}\ , \text{ with} \quad\ell_{\mu\nu}(q)\ =\ \frac{q_\mu
q_\nu}{q^2}\ .
\end{equation}
One can write two separate DRs for the transverse self-energy,
$\Pi_T$, of a massive gauge boson, which crucially depend on the
pole structure of Eq.\ (\ref{vm:oprop1}), namely
\begin{eqnarray}
\label{vm:DRphys} \Re e\bar{\Pi}_T(s) & = & \Re e\bar{\Pi}_T(M^2) +
(s-M^2)\Re e\bar{\Pi}'_T(M^2)
+\frac{(s-M^2)^2}{\pi}\nonumber\\
&&\times \mbox{P}\, \int\limits_{\{ M^2_{phys}\} }^{+\infty} ds'
\, \frac{\Im m \bar{\Pi}_T (s')}{(s'-M^2)^2(s'-s)}\, ,\\
\label{vm:DRunph} \Re e\bar{\Pi}^{(\xi_Q)}_T (s) & = & (s-M^2)\Re
e\bar{\Pi}'^{(\xi_Q)}_T(M^2) +\frac{(s-M^2)^2}{\pi}\mbox{P}\,
                              \int\limits_{\{ M^2_{unphys}\} }^{+\infty} ds'
\, \frac{\Im m \bar{\Pi}_T^{(\xi_Q)} (s')}{(s'-M^2)^2(s'-s)}\, .\nonumber\\
&&
\end{eqnarray}
In the first DR given in Eq.\ (\ref{vm:DRphys}), the real part of
$\Pi_T$, $\Re e\bar{\Pi}_T$, is determined from branch cuts induced
by physical poles, where the masses of the real on-shell particles
in the loop are collectively denoted by $\{ M^2_{phys} \}$. In what
follows we refer to such a DR as \bleu{physical} DR. Note that $\Re
e\bar{\Pi}_T$ depends only implicitly on the gauge choice. In fact,
$\Re e\bar{\Pi}_T$ can be viewed as the truncated part of the
self-energy that will survive if $\Re e{\Pi}_T$ is embedded in a
$S$-matrix element. In Eq.\ (\ref{vm:DRunph}), the dispersive part
of the two-point function depends explicitly on $\xi_Q$-dependent
unphysical thresholds, collectively denoted by $\{ M^2_{unphys}\}$,
which are induced by the longitudinal parts of the gauge propagators
contained in $\Im m\bar{\Pi}_T^{(\xi_Q)}$. Evidently, one has the
decomposition
\begin{equation}
\Im m\Pi_T (s)\ =\ \Im m\bar{\Pi}_T(s) +\Im
m\bar{\Pi}_T^{(\xi_Q)}(s)\, ,\quad \Re e\Pi_T (s)\ =\ \Re
e\bar{\Pi}_T(s) +\Re e\bar{\Pi}_T^{(\xi_Q)}(s)\, .
\end{equation}
${}$From Eq.\ (\ref{vm:oprop1}), one can now isolate that part of
the propagator that should be used in a physical DR. For $\xi_Q\not=
1$, one has
\begin{equation}
\Delta_{0\mu\nu}^{(\xi_Q)}\ \to \ U_{\mu\nu}(q)\ \equiv \
\Delta_{0\mu\nu }^{(\infty )}(q)\, .
\end{equation}
It is therefore obvious that the `physical' sector of an off-shell
transition amplitude in BFG (for $\xi_Q\not= 1$) ---or equivalently,
the part of the off-shell matrix element that satisfies a
\bleu{physical} DR--- is effectively obtained by considering all the
internal propagators in the unitary gauge ($\xi_Q\to \infty $), but
leaving the Feynman rules for the vertices in the general $\xi_Q$
gauge.

In view of a physical DR, the gauge $\xi_Q=1$ is very specific,
since the physical and unphysical poles coincide in such a case,
making them indistinguishable. At one-loop order, the results of
this gauge are found to collapse to those obtained via the
PT~\cite{Denner:1994nn}. Finally we remark in passing that, if
$\bar{\Pi}_T$ in $\xi_Q\not= 1$ is used for a definition of a
`physical' self-energy, one encounters problems with the high-energy
unitarity behaviour, even though the full $\Pi(\xi_{Q})$ is
asymptotically well-behaved. In the case of the one-loop $Z$
self-energy for example, for $\xi_{Q}\not= 1$ \cite{Denner:1994nn},
$\bar{\Pi}_T$ contains terms proportional to $q^{4}$; all such terms
eventually cancel in the entire $\Pi(\xi_{Q})$ against the part that
contains the unphysical poles. Incidentally, it is interesting to
notice that the recovery of the correct asymptotic behaviour is the
more delayed, {\em i.e.}, it happens for larger values of $q^{2}$,
the larger the value of $\xi_{Q}$. However, if one was to resum only
the $\bar{\Pi}_T$ part, the terms proportional to $q^{4}$ would
survive, leading to bad high energy behaviour. If, on the other
hand, one had resummed the full $\Pi(\xi_{Q})$, then one would have
introduced unphysical poles, as explained above.

\subsection {Unitarity and gauge invariance} \indent

In this section, we will briefly discuss the basic
field-theoretical consequences resulting from the unitarity of the
$S$-matrix theory, and establish its connection with gauge
invariance. In addition to the requirement of explicit gauge
invariance, the necessary conditions derived from unitarity will
constitute our guiding principle to analytically continue
$n$-point correlation functions in the off-shell region.
Furthermore, we arrive at the important conclusion that the
resummed self-energies, in addition to being GFP independent, must
also be ``unitary'', in the sense that they do not spoil unitarity
when embedded in an $S$-matrix element.

The $T$-matrix element of a reaction $i\to f$ is defined via the
relation
\begin{equation}
\langle f | S | i \rangle\ =\ \delta_{fi}\ +\ i(2\pi )^4
\delta^{(4)}(P_f - P_i)\langle f | T|i\rangle ,
\end{equation}
where $P_i$ ($P_f$) is the sum of all initial (final) momenta of the
$|i\rangle$ ($| f \rangle$) state. Furthermore, imposing the
unitarity relation $S^\dagger S = 1$, consequence of the
conservation of the probability,  leads to the optical theorem:
\begin{equation}
\langle f|T|i\rangle - \langle i |T|f\rangle^*\ =\ i\sum_{i'} (2\pi
)^4\delta^{(4)}(P_{i'} - P_i)\langle i' | T | f \rangle^* \langle i'
| T | i \rangle. \label{vm:optical}
\end{equation}
In Eq.~(\ref{vm:optical}), the sum $\sum_{i'}$ should be understood
to be over the entire phase space and spins of all possible on-shell
intermediate particles $i'$. A corollary of this theorem is obtained
if $i=f$. In this particular case, we have
\begin{equation}
2\Im m \langle i|T|i\rangle\ =\sum_f (2\pi )^4\delta^{(4)}(P_f -
P_i) |\langle f | T|i \rangle |^2. \label{vm:absorptive}
\end{equation}
In the conventional $S$-matrix theory with stable particles,
Eqs~(\ref{vm:optical}) and (\ref{vm:absorptive}) hold also
perturbatively. To be precise, if one expands the transition
$T=T^{(1)}+T^{(2)}+\cdots + T^{(n)} + \cdots$, to a given order $n$,
one has
\begin{equation}
T^{(n)}_{fi}-T^{(n)*}_{if}\ =\ i\sum_{i'} (2\pi )^4
\delta^{(4)}(P_{i'}-P_i) \sum\limits_{k=1}^{n-1} T^{(k)*}_{i'f}
T^{(n-k)}_{i'i}.\label{vm:optpert}
\end{equation}
There are two important conclusions that can be drawn from Eq.\
(\ref{vm:optpert}). First, the anti-hermitian part of the LHS of
Eq.\ (\ref{vm:optpert}) contains, in general, would-be Goldstone
bosons or ghost fields \cite{FP}. Such contributions manifest
themselves as Landau singularities at unphysical points, {\em e.g.},
$q^2=\xi_Q M_W^2$ for a $W$ propagator in a general BFG. However,
unitarity requires that these unphysical contributions should
vanish, as can be read off from the RHS of Eq.\ (\ref{vm:optpert}).
Second, the RHS explicitly shows the connection between gauge
invariance and unitarity at the quantum loop level. To lowest order
for example, the RHS consists of the product of GFP independent
on-shell tree amplitudes, thus enforcing the gauge-invariance of the
imaginary part of the one-loop amplitude on the LHS.

The above powerful constraints imposed by unitarity will be in
effect as long as one computes \bleu{full} amplitudes to a finite
order in perturbation theory. However, for resummation purposes, a
certain sub-amplitude, {\em i.e.}, a part of the full amplitude,
must be singled out and subsequently undergo a Dyson summation,
while the rest of the $S$-matrix is computed to a finite order $n$.
Therefore, if the resummed amplitude contains gauge artifacts and/or
unphysical thresholds, the cancellations imposed by Eq.\
(\ref{vm:optpert}) will only operate up to order $n$, introducing
unphysical contributions of order $n+1$ or higher. To avoid the
contamination of the physical amplitudes by such unphysical
artifacts, we impose the following two requirements on the effective
Green's functions, when one attempts to continue them analytically
in the off-shell region for the purpose of resummation:

\begin{itemize}

\item[(i)] The off-shell $n$-point correlation functions ought to
be derivable from or embeddable into $S$-matrix elements.

\item[ (ii)] The off-shell Green's functions should not display
unphysical thresholds induced by unphysical Landau singularities,
as has been described above.

\end{itemize}

Even though property (i) is automatic for Green's functions
generated by the functional differentiation of the conventional
path-integral functional, in general the off-shell amplitudes so
obtained fail to satisfy property (ii). In the PT framework instead,
both conditions are satisfied: Effective Green's functions are
directly derived from the $S$-matrix amplitudes (so condition (i) is
satisfied by construction) and contain only physical thresholds, so
that unitarity is not explicitly violated \cite{JP&AP}.

In our discussion of unitarity at one-loop, we will make extensive
use of the following two-body Lorentz-invariant phase-space (LIPS)
integrals: The scalar integral
\begin{eqnarray}
\int dX_{LIPS} &=& \frac{1}{(2\pi)^2}\, \int d^{4}k_1 \int
d^{4}k_2\, \delta_+ (k^2_1-m^2_1)\delta_+ (k^2_2-m^2_2)
\delta^{(4)}(q-k_1-k_2)
\nonumber\\
&=& \theta(q^0)\theta [q^2-{(m_1+m_2)}^2]\, \frac{1}{8\pi\, q^2}\,
\lambda^{1/2} (q^2,m_1^2,m_2^2)\, , \label{vm:LIPS1}
\end{eqnarray}
where $\lambda (x,y,z)= (x-y-z)^2-4yz$ and
$\delta_+(k^2-m^2)\equiv \theta (k^0)\delta(k^2-m^2)$, and the
tensor integral:
\begin{eqnarray}
\int dX_{LIPS} {(k_1-k_2)}_{\mu}{(k_1-k_2)}_{\nu} &=& \Big\{
\frac{\lambda (q^2,m_1^2,m_2^2)}{3q^2}\, t_{\mu\nu}(q) +\,
\Big[ \frac{\lambda (q^2,m_1^2,m_2^2)}{q^2} - q^2\nonumber\\
&&+ 2 (m^2_1+m^2_2) \Big]\, \ell_{\mu\nu}(q)\, \Big\} \times \int
dX_{LIPS}\, . \label{vm:LIPS2}
\end{eqnarray}

\section{The absorptive pinch technique construction}
\label{vm:sec:absorptivePT}
\subsection {Forward scattering in QCD}

In this section, we show that a self-consistent picture may be
obtained by resorting to such fundamental properties of the
$S$-matrix as unitarity and analyticity, using as additional input
only elementary Ward identities (EWIs) for tree-level, on-shell
processes, and tree-level vertices and propagators.  It is important
to emphasize that the gauge-fixing parameter (GFP) independence of
the results emerges \bleu{automatically} from the previous
considerations.

We begin from the right-hand-side (RHS) of the optical relation
given in Eq.\ (\ref{vm:absorptive}). The RHS involves on-shell
physical processes, which satisfy the EWIs. It turns out that the
full exploitation of those EWIs leads unambiguously to a
decomposition of the tree-level amplitude into propagator-,
vertex- and box-like structures. The propagator-like structure
corresponds to the imaginary part of the effective propagator
under construction. By imposing the additional requirement that
the effective propagator be an analytic function of $q^2$ one
arrives at a  dispersion relation (DR), which, up to
renormalization-scheme choices, leads to a unique result for the
real part.

\begin{figure}[htb]
\begin{center}
\includegraphics[width=12cm]{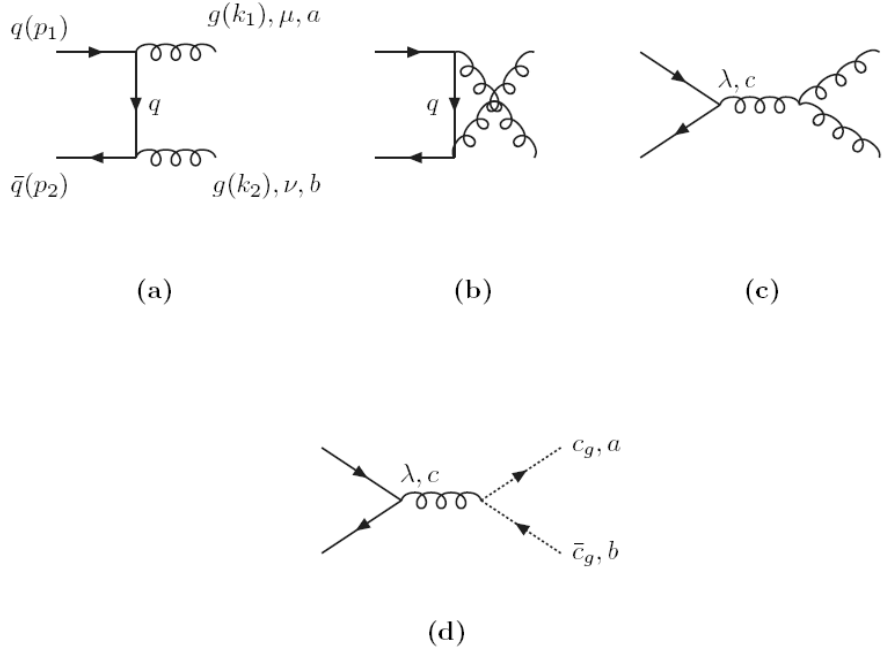}
\end{center}
\caption{\label{vm:fig:qqscat}Diagrams (a)--(c) contribute to ${\cal
T}^{ab}_{\mu\nu}$, and diagram (d) to ${\cal S}^{ab}$.}
\end{figure}

Consider the forward scattering process $q {\bar q}\rightarrow
q{\bar q}$. From the optical theorem, we then have
\begin{equation}
2\Im m \langle q\bar{q}|T|q\bar{q}\rangle\ =\left( \frac{1}{2}
\right)\, \int dX_{LIPS}\, \langle q\bar{q}|T|gg\rangle \langle
gg|T|q\bar{q}\rangle^{*}\, . \label{vm:OTgg}
\end{equation}
In Eq.\ (\ref{vm:OTgg}), the statistical factor 1/2 in parentheses
arises from the fact that the final on-shell gluons should be
considered as identical particles in the total rate. We consider
only physical gluon as intermediate states. The inclusion of
quarks will lead to the contribution of the quark loop in the
self-energy. We now set ${\cal M}=\langle
q\bar{q}|T|q\bar{q}\rangle $ and ${\cal T}=\langle
q\bar{q}|T|gg\rangle$, and focus on the RHS of Eq.\
(\ref{vm:OTgg}). Diagrammatically, the amplitude ${\cal T}$
consists of two distinct parts: $t$ and $u$-channel graphs that
contain an internal quark propagator, ${{\cal
T}_{t}}^{ab}_{\mu\nu}$, as shown in Figs.\ \ref{vm:fig:qqscat}(a)
and \ref{vm:fig:qqscat}(b), and an $s$-channel amplitude, ${{\cal
T}_{s}}^{ab}_{\mu\nu}$, which is given in Fig.\
\ref{vm:fig:qqscat}(c). The subscript ``$s$'' and ``$t$'' refers
to the corresponding Mandelstam variables, {\em i.e.}\ $s=q^2=
(p_1+p_2)^2=(k_1+k_2)^2$, and $t=(p_1-k_1)^2=(p_2-k_2)^2$.
Defining the quark current
\begin{equation}
V_{\rho}^{c}\ =\ g\bar{v}(p_2)\, \frac{\lambda^c}{2}\gamma_{\rho}\,
u(p_1)\, ,
\end{equation}
we have that
\begin{equation}
{\cal T}^{ab}_{\mu\nu}={{\cal T}_{s}}^{ab}_{\mu\nu} (\xi )+ {{\cal
T}_{t}}^{ab}_{\mu\nu}\, , 
\end{equation}
with
\begin{eqnarray}
{{\cal T}_{s}}^{ab}_{\mu\nu}(\xi )& =& -gf^{abc}\, \Delta^{(\xi
),\rho\lambda}_0(q) \Gamma_{\lambda\mu\nu}(q,-k_1,-k_2)\,
V_{\rho}^{c}\, ,
\label{vm:Ts}\\
{{\cal T}_{t}}^{ab}_{\mu\nu} &=& -ig^2\bar{v}(p_2)\Big( \,
\frac{\lambda^b}{2}\gamma^{\nu}\, \frac{1}{\not\! p_1-\not\! k_1 -
m} \, \frac{\lambda^a}{2}\gamma^{\mu}\ +\
\frac{\lambda^a}{2}\gamma^{\mu}\, \frac{1}{\not\! p_1-\not\!
k_2-m}\, \gamma^{\nu}\frac{\lambda^b}{2}\, \Big)u(p_1)\, ,\qquad
\label{vm:Tt}
\end{eqnarray}
where
\begin{equation}
\Gamma_{\lambda\mu\nu}(q,-k_1,-k_2)\ =\
(k_1-k_2)_{\lambda}g_{\mu\nu}\, +\, (q+k_2)_{\mu}g_{\lambda\nu}\,
-\, (q+k_1)_{\nu}g_{\lambda\nu}\, . \label{vm:3GV}
\end{equation}

Notice that ${\cal T}_{s}$ depends explicitly on the GFP $\xi$,
through the tree-level gluon propagator
$\Delta^{(\xi)}_{0\mu\nu}(q)$, whereas ${\cal T}_{t}$ does not.
The explicit expression of $\Delta^{(\xi )}_{0\mu\nu}(q)$ depends
on the specific gauge fixing procedure chosen.  In addition, we
define the quantities\footnote{Note that ${\cal S}^{ab}$ is a
ghost-like amplitude.} ${\cal S}^{ab}$ and ${\cal R}^{ab}_{\mu}$
as follows:
\begin{eqnarray}
{\cal S}^{ab}&=&gf^{abc}\, \frac{k^\sigma_1}{q^2}\,
V_{\sigma}^{c} 
=-gf^{abc}\, \frac{k^\sigma_2}{q^2}\, V_{\sigma}^{c} \label{vm:S}
\end{eqnarray}
and
\begin{equation}
{\cal R}_{\mu}^{ab}\ =\ gf^{abc}\, V_{\mu}^{c}\, .
\end{equation}
Clearly,
\begin{equation}
k_1^{\sigma}{\cal R}_{\sigma}^{ab}\ =\ -k_2^{\sigma}{\cal
R}_{\sigma}^{ab}\ =\ q^2{\cal S}^{ab}. \label{vm:D2}
\end{equation}
We then have
\begin{eqnarray}
\Im m {\cal M}&=& \frac{1}{4}\, {\cal T}^{ab}_{\mu\nu}\,
P^{\mu\sigma} (k_1,\eta_1)\, P^{\nu\lambda}(k_2,\eta_2)\, {\cal
T}^{ab*}_{\sigma\lambda}
\nonumber\\
&=& \frac{1}{4}\Big[ {{\cal T}_{s}}^{ab}_{\mu\nu}(\xi)+ {{\cal
T}_{t}}^{ab}_{\mu\nu}\Big]\, P^{\mu\sigma}(k_1,\eta_1)\,
P^{\nu\lambda}(k_2,\eta_2)\, \Big[ {{\cal
T}_{s}}^{ab*}_{\sigma\lambda}(\xi) +{{\cal
T}_{t}}^{ab*}_{\sigma\lambda}\Big], \label{vm:MM}
\end{eqnarray}
where the polarization tensor $P^{\mu\nu}(k,\eta )$ is given by
\begin{equation}
\sum_{\text{Phys.}}\epsilon_\mu^a(k)\epsilon_\nu^a(k) =
P_{\mu\nu}(k,\eta )\ =\ -g_{\mu\nu}+ \frac{\eta_{\mu}k_{\nu}
+\eta_{\nu}k_{\mu} }{\eta k} + \eta^2 \frac{k_{\mu}k_{\nu}}{{(\eta
k)}^2}\, . \label{vm:ophotPol}
\end{equation}
Moreover, we have that on-shell, {\em i.e.}, for $k^{2}=0$,
$k^{\mu}P_{\mu\nu}=0$. By virtue of  this last property, we see
immediately that if we write the three-gluon vertex of Eq.\
(\ref{vm:3GV}) in the form
\begin{eqnarray}
\Gamma_{\lambda\mu\nu} (q,-k_1,-k_2) &=&
[(k_1-k_2)_{\lambda}g_{\mu\nu}+
2q_{\mu}g_{\lambda\nu}-2q_{\nu}g_{\lambda\mu}]\
+\ (-k_{1\mu}g_{\lambda\nu}+k_{2\nu}g_{\lambda\mu}) \nonumber\\
&=& \Gamma^F_{\lambda\mu\nu}(q,-k_1,-k_2)\ +\
\Gamma^P_{\lambda\mu\nu}(q,-k_1,-k_2)\, , 
\end{eqnarray}
the term $\Gamma^P_{\rho\mu\nu}$ dies after hitting the polarization
vectors $P_{\mu\sigma}(k_1,\eta_1)$ and
$P_{\nu\lambda}(k_2,\eta_2)$.  Therefore, if we denote by ${\cal
T}_{s}^{F}(\xi)$ the part of ${\cal T}_{s}$ which survives, Eq.\
(\ref{vm:MM}) becomes
\begin{equation}
\Im m {\cal M}\ =\ \frac{1}{4}\, \big[ {\cal T}_{s}^{F}(\xi)+{\cal
T}_{t}\big]^{ab}_{\mu\nu}\, P^{\mu\sigma}(k_1,\eta_1 )\,
P^{\nu\lambda}(k_2,\eta_2  )\, \big[ {\cal T}_{s}^{F}(\xi) +{\cal
T}_{t} \big]^{ab*}_{\sigma\lambda}\, . \label{vm:MM22}
\end{equation}
The next step is to verify that any dependence on the GFP inside the
propagator $\Delta^{(\xi)}_{0\mu\nu}(q)$ of the off-shell gluon will
disappear. This is indeed so, because the longitudinal parts of
$\Delta_{0\mu\nu}$ either vanish because the external quark current
is conserved, or because they trigger the following EWI:
\begin{equation}
q^{\mu}\Gamma^{F}_{\mu\alpha\beta}(q, -k_1, -k_2)\ =\ (k_1^2\ -\
k_2^2)g_{\alpha\beta}\, , \label{vm:FWI}
\end{equation}
which vanishes on shell.  This last EWI is crucial, because in
general, current conservation alone is not sufficient to guarantee
the GFP independence of the final answer. In the covariant gauges
for example, the gauge fixing term is proportional to
$q^{\mu}q^{\nu}$; current conservation kills such a term. But if we
had chosen an axial gauge instead, {\em i.e.}
\begin{equation}
\label{vm:Deleta} \Delta^{(\tilde{\eta})}_{0\mu\nu}(q)\ =\
\frac{iP_{\mu\nu} (q,\tilde{\eta})}{q^2}\, ,
\end{equation}
where $\tilde{\eta} \neq \eta$ in general, then only the term
${\tilde{\eta}_{\nu}}q_{\mu}$ vanishes because of current
conservation, whereas the term ${\tilde{\eta}_{\mu}}q_{\nu}$ can
only disappear if Eq.\ (\ref{vm:FWI}) holds.  So, Eq.\
(\ref{vm:MM22}) becomes
\begin{equation}
\Im m {\cal M}\ =\ \frac{1}{4} ({\cal T}_{s}^{F}+{\cal
T}_{t})^{ab}_{\mu\nu}\, P^{\mu\sigma}(k_1,\eta_1)\,
P^{\nu\lambda}(k_2,\eta_2)\, ({\cal T}_{s}^{F}+{\cal
T}_{t})^{ab*}_{\sigma\lambda}\, , \label{vm:MM2}
\end{equation}
where the GFP-\bleu{independent} quantity ${\cal T}_{s}^{F}$ is
given by
\begin{equation}
{{\cal T}_{s}}^{F,ab}_{\mu\nu}\ =\ -gf^{abc}\,
\frac{g^{\rho\lambda}}{q^2}\, \,
\Gamma^{F}_{\lambda\mu\nu}(q,-k_1,-k_2)\, V_{\rho}^{c}\, .
\label{vm:TsF}
\end{equation}
Next, we want to show that the dependence on $\eta_{\mu}$ and
$\eta^2$ stemming from the polarization vectors disappears.  Using
the on shell conditions $k_1^2=k_2^2=0$, we can easily verify the
following EWIs:
\begin{eqnarray}
k_1^{\mu}{{\cal T}_{s}}^{F,ab}_{\mu\nu} & = & 2 k_{2\nu}{\cal
S}^{ab}\, -\, {\cal R}_{\nu}^{ab} \, ,
\label{vm:w1}\\
k_2^\nu {{\cal T}_{s}}^{F,ab}_{\mu\nu}  & = &  2 k_{1\mu}{\cal
S}^{ab}
\, +\, {\cal R}_{\mu}^{ab}  \, ,\label{vm:w2}\\
k_1^{\mu}{{\cal T}_{t}}^{ab}_{\mu\nu} & = & {\cal R}_{\nu}^{ab} \, ,
\label{vm:w3}\\
k_2^{\nu}{{\cal T}_{t}}^{ab}_{\mu\nu} & = & -{\cal R}_{\mu}^{ab}
\label{vm:w4}\, ,
\end{eqnarray}
from which we have that
\begin{eqnarray}
k_1^{\mu}k_2^{\nu}{{\cal T}_{s}}^{F,ab}_{\mu\nu}& = & q^2{\cal
S}^{ab}
\label{vm:w5}\, , \\
k_1^{\mu}k_2^{\nu}{{\cal T}_{t}}^{ab}_{\mu\nu} &=& -q^2{\cal
S}^{ab}\, . \label{vm:w6}
\end{eqnarray}
Using the above EWIs, it is now easy to check that indeed, all
dependence on both $\eta_{\mu}$ and $\eta^2$ cancels in Eq.\
(\ref{vm:MM2}), as it should, and we are finally left with (omitting
the fully contracted colour and Lorentz indices):
\begin{eqnarray}
\Im m {\cal M} &=& \frac{1}{4}\, \Big[ \Big( {\cal T}_{s}^{F}{{\cal
T}_{s}^{F}}^{*} -8 {\cal S}{\cal S}^{*}\Big) + \Big( {\cal
T}_{s}^{F}{\cal T}_{t}^{*} + {{\cal T}_{s}^{F}}^{*}
{\cal T}_{t} \Big) + {\cal T}_{t}{\cal T}_{t}^{*} \Big]\nonumber\\
&=& \Im m \widehat{{\cal M}}_1+ \Im m \widehat{{\cal M}}_2+ \Im m
\widehat{{\cal M}}_{3} \, . \label{vm:MM3}
\end{eqnarray}
The first part is the genuine propagator-like piece (sum of a gluon
and ghost parts), the second is the vertex, and the third the box.
Employing the fact that

\begin{equation}
\Gamma^{F}_{\rho\mu\nu}\Gamma^{F,\mu\nu}_{\lambda} \ =\
-8q^2t_{\rho\lambda}(q) + 4{(k_1-k_2)}_{\rho}{(k_1-k_2)}_{\lambda}
\end{equation}
and
\begin{eqnarray}
{\cal S}{\cal S }^{*} &=& g^2\, N\, V^c_\rho \,
\frac{k_1^{\rho}k_1^{\lambda}}{(q^2)^2}\, V^{c}_{\lambda} \nonumber\\
&=& \frac{g^2}{4}\, N\, V^c_{\rho}\, \frac{(k_1-k_2)^\rho
(k_1-k_2)^\lambda }{(q^2)^2}\, V^{c}_{\lambda}\, ,
\end{eqnarray}
where $N$ is the eigenvalue of the Casimir operator in the adjoint
representation for SU$(N)$, we obtain for $\Im m \widehat{{\cal
M}}_1$
\begin{equation}
\Im m \widehat{{\cal M}}_1\ =\ \frac{g^2}{2}\, N V^{c}_{\mu}\,
\frac{1}{q^2}\, \Big[ -4q^2t^{\mu\nu}(q)\, +\,
{(k_1-k_2)}^{\mu}{(k_1-k_2)}^{\nu}\Big]\, \frac{1}{q^2}\,
V^{c}_{\nu}\, .
\end{equation}
This last expression must be integrated over the available phase
space. With the help of Eqs.\ (\ref{vm:LIPS1}) and (\ref{vm:LIPS2}),
we arrive at the final expression
\begin{equation}
\Im m \widehat{{\cal M}}_1\ =\ V^c_\mu\, \frac{1}{q^2}\, \Im m
\widehat{\Pi}^{\mu\nu}(q)\, \frac{1}{q^2}\, V^c_\nu\, ,
\end{equation}
with
\begin{equation}
\label{vm:IMQCD} \Im m \widehat{\Pi}_{\mu\nu}(q)\ =\ -\,
\frac{\alpha_s}{4}\, \frac{11N}{3}\, q^2 t_{\mu\nu}(q)\, ,
\end{equation}
and $\alpha_s=g^2/(4\pi)$.

Before we proceed, we make the following remark. It is well-known
that the vanishing of the longitudinal part of the gluon self-energy
is an important consequence of gauge invariance. One might naively
expect that even if a non-vanishing longitudinal part had been
induced by some contributions which do not respect gauge invariance,
it would not have contributed to physical processes, since the gluon
self-energy couples to conserved fermionic currents, thus projecting
out only the transverse degrees of the gluon vacuum polarization.
However, this expectation is not true in general. Indeed, if one
uses, for example, the tree-level gluon propagator in the axial
gauge, as given in Eq.~\eqref{vm:Deleta}, then there will be
residual $\eta$-dependent terms induced by the longitudinal
component of the gluon vacuum polarization, which would not vanish,
despite the fact that the external quark currents are conserved.
Such terms are obviously gauge dependent. Evidently, projecting out
only the transverse parts of Green's functions will not necessarily
render them gauge invariant.

The vacuum polarization of the gluon within the PT is given by
\begin{equation}
\label{vm:opTgg} \widehat{\Pi}_{\mu\nu}(q)\ =\
\frac{\alpha_s}{4\pi}\, \frac{11N}{3}\, t_{\mu\nu}(q)\, q^2\,
\Big[\, \ln\Big(\frac{q^2}{\mu^2}\Big)\, +\, C_{UV}\, \Big]\, .
\end{equation}
Here, $C_{UV}=1/\epsilon -\gamma + \ln 4\pi + C$, with $C$ being
some constant and $\mu$ is a subtraction point. In Eq.\
(\ref{vm:opTgg}), it is interesting to notice that a change of
$\mu^2\to \mu '^2$ gives rise to a variation of the constant $C$ by
an amount $C'-C=\ln \mu'^2/\mu^2$.  Thus, a general $\mu$-scheme
renormalization yields
\begin{eqnarray}
\label{vm:RPTgg} \widehat{\Pi}_T^R (s) &=& \widehat{\Pi}_T (s)\, -\,
(s-\mu^2)\Re e
\widehat{\Pi}_T'(\mu^2)\, -\, \Re e\widehat{\Pi}_T(\mu^2 ) \nonumber\\
&=& \frac{\alpha_s}{4\pi}\, \frac{11N}{3}\, s\, \Big[
\ln\Big(\frac{s}{\mu^2}\Big)\, -\, 1\, +\, \frac{\mu^2}{s}\, \Big]\,
.
\end{eqnarray}
${}$From Eq.\ (\ref{vm:DR3}), one can readily see that $\Re e
\widehat{\Pi}^R_T(s)$ can be calculated by the following double
subtracted DR:
\begin{equation}
\label{vm:QCDvm:DR} \Re e \widehat{\Pi}^R_T(s)\ =\
\frac{(s-\mu^2)^2}{\pi}\ \mbox{P}\, \int\limits_0^\infty\, ds'
\frac{\Im m \widehat{\Pi}_T(s')}{(s'-\mu^2)^2 (s'-s)}\, .
\end{equation}
Inserting Eq.\ (\ref{vm:IMQCD}) into Eq.\ (\ref{vm:QCDvm:DR}), it is
not difficult to show that it leads to the result given in Eq.\
(\ref{vm:RPTgg}), a fact that demonstrates the analytic power of the
DR.

It is important to emphasize that the above derivation rigorously
proves the GFP independence of the one-loop PT effective Green's
functions, for \bleu{every} gauge fixing procedure. Indeed, in our
derivation, we have solely relied on the RHS of the OT, which we
have rearranged in a well-defined way, \bleu{after} having
explicitly demonstrated its GFP-independence. The proof of the
GFP-independence of the RHS presented here is, of course, expected
on physical grounds, since it only relies on the use of EWIs,
triggered by the longitudinal parts of the gluon tree-level
propagators. Note that the tree-level tri-gluon coupling,
$\Gamma_{\lambda\mu\nu}$, is uniquely given by Eq.\
(\ref{vm:3GV}). Since the GFP-dependence is carried entirely by
the longitudinal parts of the gluon tree-level propagator in {\em
any} gauge-fixing scheme whereas the $g^{\mu\nu}$ part is
GFP-independent and universal, the proof presented here is
generally true. Obviously, the final step of reconstructing the
real part from the imaginary by means of a DR does not introduce
any gauge-dependences.

\subsection{The QCD analysis from BRS considerations} \indent

In this section, we will show how we can obtain the same answer by
resorting only to the EWIs that one obtains as a direct consequence
of the BRS symmetry of the quantum Lagrangian.

If we consider ${\cal T}_{\mu\nu}^{ab}$ as before, it is easy to
show that it satisfies the following BRS identities \cite{ChengLi}:
\begin{eqnarray}
k^{\mu}_1{\cal T}_{\mu\nu}^{ab} &=& k_{2\nu}{\cal S}^{ab}\, ,\nonumber\\
k^{\nu}_2{\cal T}_{\mu\nu}^{ab} &=& k_{1\mu}{\cal S}^{ab}\, ,  \nonumber\\
k^{\mu}_1k^{\nu}_2{\cal T}_{\mu\nu}^{ab} &=& 0\, , \label{vm:BRSg}
\end{eqnarray}
where $S^{ab}$ is the ghost amplitude shown in Fig.\
\ref{vm:fig:qqscat}(d); its closed form is given in Eq.\
(\ref{vm:S}).

Notice that the BRS identities of Eq.\ (\ref{vm:BRSg}) are different
from those listed in Eqs. (\ref{vm:w1})--(\ref{vm:w6}), because  the
term $\Gamma_{\mu\nu\rho}^{P}$ had been removed in the latter case.
Here, we follow a different sequence and do not kill the term
$\Gamma_{\mu\nu\rho}^{P}$; instead, we will exploit the \bleu{exact}
BRS identities from the very beginning.

We start again with the expression for $\Im m {\cal M}$ given in
Eq.\ (\ref{vm:MM}). First of all, it is easy to verify again that
the dependence on the GFP of the off-shell gluon vanishes. This is
so because of the tree-level EWI, involving the \bleu{full} vertex
$\Gamma_{\mu\nu\rho}$,
\begin{equation}
q^{\lambda}\Gamma_{\lambda\mu\nu}(q,-k_1,-k_2)\ =\ k^2_2\,
t_{\mu\nu}(k_2 )\ -\ k^2_1\, t_{\mu\nu}(k_1 )\, .
\end{equation}
The RHS vanishes after contracting with the polarization vectors,
and employing the on-shell condition $k^2_1=k^2_2=0$.  Again, by
virtue of the BRS identities and the on-shell condition
$k_1^2=k_2^2=0$, the dependence of $\Im m {\cal M}$ on the
parameters $\eta_{\mu}$ and $\eta^2$ cancels, and we eventually
obtain
\begin{eqnarray}
\Im m {\cal M} &=& \frac{1}{4}\,  {\cal T}_{\mu\nu}\,
P^{\mu\rho}(k_1,\eta_1)\, P^{\nu\sigma}(k_2,\eta_2)\, {\cal
T}_{\rho\sigma}^*
\nonumber\\
&=& \frac{1}{4}\, \Big( {\cal T}^{\mu\nu} {\cal T}_{\mu\nu}^{*}\ -\
2 {\cal S}{\cal S}^{*}\Big) \nonumber\\
&=& \frac{1}{4}\, \Big[({\cal T}_{s}^{F}+{\cal T}_{s}^{P}+{\cal
T}_{t})^{\mu\nu} ({\cal T}_{s}^{F}+{\cal T}_{s}^{P}+{\cal
T}_{t})_{\mu\nu}^{*}\ -\ 2 {\cal S}{\cal S}^{*}\Big]\, ,
\label{vm:GIFA}
\end{eqnarray}
where
\begin{equation}
{{\cal T}_{s}}^{P,ab}_{\mu\nu}= -gf^{abc}\, \frac{g^{\rho
\lambda}}{q^2}\, \Gamma^{P}_{\lambda\mu\nu}(q,-k_1,-k_2)\, V_{\rho
}^{c}\, . \label{vm:TsP}
\end{equation}

At this point, one  must recognize that due to the four-momenta of
the trilinear vertex $\Gamma^{P}$ inside ${\cal T}_{s}^{P}$, one can
further trigger the EWIs, exactly as one did in order to derive from
Eq.\ (\ref{vm:MM}) the last step of Eq.\ (\ref{vm:GIFA}). In fact,
only the process-independent terms contained in $\Im m {\cal M}$
will be projected out on account of the BRS identities of Eq.\
(\ref{vm:BRSg}). It is important to emphasize that ${\cal
T}_{s}^{F}$ and ${\cal T}_{t}$ do not contain any pinching momenta.
This is particular to this example, where we have only two gluons as
final states, but is not true for more gluons.  To further exploit
the EWIs derived from BRS symmetries, we re-write the RHS of Eq.\
(\ref{vm:GIFA}) in the following way (we omit the fully contracted
Lorentz indices):
\begin{eqnarray}
\Im m {\cal M} &=& \frac{1}{4}\, \Big[({\cal T}_{t}+{\cal
T}^{P}_{s}+{\cal T}^{F}_{s}) ({\cal T}_{t}+{\cal T}^{P}_{s}+{\cal
T}^{F}_{s})^{*}\ -\ 2 {\cal S}{\cal S}^{*} \Big]
\nonumber\\
&=& \frac{1}{4}\, \Big[({\cal T}^{F}_{s}{{\cal T}^{F}_{s}}^{*}
-{\cal T}^{P}_{s}{{\cal T}^{P}_{s}}^{*} +{\cal T}^{P}_{s}{\cal
T}^{*}+{\cal T}{{\cal T}^{P}_{s}}^{*} - 2 {\cal S} {\cal S}^{*}) +(
{\cal T}_{t} {{\cal T}^{F}_{s}}^{*} +{\cal T}^{F}_{s}{\cal
T}_{t}^{*} ) +{\cal T}_{t}{\cal T}_{t}^{*}\Big]
\nonumber\\
&=& \Im m \widehat{{\cal M}}_1+\Im m \widehat{{\cal M}}_2+ \Im m
\widehat{{\cal M}}_{3}\, . \label{vm:makrynari}
\end{eqnarray}
In Eq.\ (\ref{vm:makrynari}), the reader may recognize the
rearrangement characteristic of the ``intrinsic'' PT, presented in
Sec.~\ref{vm:sec:intrinsic}.

Inserting the explicit form of ${\cal T}^{P}_s$ given in Eq.\
(\ref{vm:TsP}) into Eq.\ (\ref{vm:makrynari}) and using the BRS
identities,
\begin{eqnarray}
\label{vm:TPsid}
{\cal T}^{P}_{s} {\cal T}^{*} &=& -2 {\cal S}{\cal S}^{*}\, ,\nonumber\\
{\cal T}^{P}_{s}{\cal T}^{P*}_{s} &=& 2 {\cal S}{\cal S}^{*}\, ,
\end{eqnarray}
we obtain
\begin{eqnarray}
\Im m \widehat{{\cal M}}_1 &= & \frac{1}{4}\, \Big( {\cal
T}^{F}_{s}{{\cal T}^{F}_{s}}^{*}-{\cal T}^{P}_{s}{{\cal
T}^{P}_{s}}^{*} +{\cal T}^{P}_{s}{\cal T}^{*}+{{\cal
T}^{P}_{s}}^{*}{\cal T}
-2{\cal S}{\cal S}^{*}\Big)\nonumber\\
&=& \frac{1}{4}\, \Big( {\cal T}^{F}_{s}{{\cal
T}^{F}_{s}}^{*}-8{\cal S}{\cal S}^{*}\Big)\, , \label{vm:Final}
\end{eqnarray}
which is the same result found in the previous section, {\em i.e.},
Eq.\ (\ref{vm:MM3}).

An interesting by-product of the above analysis is that one is able
to show the independence of the PT results  of the number of the
external fermionic currents. Indeed, the BRS identities in
Eqs~\eqref{vm:BRSg}, as well as those given in Eq.~\eqref{vm:TPsid},
will still hold for any transition amplitude of $n$-fermionic
currents to two gluons.  By analogy, one can decompose the
transition amplitude into ${\cal T}_t$ and ${\cal T}_s$ structures.
Similarly, the form of the sub-structures ${\cal T}^F_s$ and ${\cal
T}^P_s$ will then change accordingly. In fact, the only modification
will be that the vector current, $V^c_\rho$, contained in Eqs.\
(\ref{vm:TsF}) and (\ref{vm:TsP}) will now represent the transition
of one gluon to $n$-fermionic currents. Making use of the
``intrinsic'' PT, one then obtains the result given in
Eq.~\eqref{vm:Final}. Hence, we can conclude that the PT does not
depend on the number of the external fermionic currents attached to
gluons.

\section{Conclusion}
We presented in this notes two versions of the pinch technique.
The S-matrix pinch technique where the idea is to start with
something we know to be gauge invariant to extract Green's
functions with physical properties. But resuming diagrams is a
quite tedious task when the number of graphs increases. In the
intrinsic version of the pinch technique, we let the pinch part of
the vertex $\Gamma^P$ acting on the full vertex $\Gamma$. The Ward
identities triggered generate terms proportional to the incoming
momenta $q_i^2$. We simply drop these terms, cancelled in the
S-matrix pinch technique by the pinch part coming from the others
diagrams. These algorithm becomes lengthy as the number of loop
increases. Fortunately, a correspondence was found with the
background field method computed in the Feynman gauge. Finally, We
review the absorptive pinch technique construction, how pinching
at tree-level generates unitarity cuts of the one-loop PT Green's
functions.

\section*{Acknowledgements}
The author would like to thank Joannis Papavassiliou for informative
and helpful discussions on the pinch technique and Alice Dechambre
for her comments about this manuscript. I would like also to thank
the Solvay Institutes for the Modave Summer Schools in Mathematical
Physics and the IISN for financial support.

\appendix

\section{Ward identities}\label{vm:sec:WI}
In classical mechanics, each symmetry provides a conserved current
given by the Noether theorem. The quantum analogy to these
conserved currents are constraints on the generating functional,
and hence on the Green's functions of the theory. When the
symmetry is the gauge invariance, the relations are called the
Ward identities, expressing that the divergences of Green's
functions vanish up to contact terms.

\subsection*{Ward identities in the background field method}

Thorough this lecture notes we speak about the Ward identities. They
are derived in QED by performing a particular change of variables (a
gauge transformation) in the generating functional. In the BFM, they
read in term of the effective action
\begin{equation}\label{vm:WI_QED}\boxed{
    \partial_\mu\frac{\delta\tilde\Gamma}{\delta A^a_\mu(x)} +
    gf^{abc}A^b_\mu\frac{\delta\tilde\Gamma}{\delta A^c_\mu(x)} +
    ig\bm T^a\psi\frac{\delta\tilde\Gamma}{\delta\psi(x)} -
    ig\bm T^a\bar\psi\frac{\delta\tilde\Gamma}{\delta\bar\psi(x)} =
    0.}
\end{equation}
This relation, given by $\delta\tilde\Gamma=0$, expresses the gauge
invariance of the theory and imposes constraints on the irreducible
Green's functions (self-energies, vertex,...).

Now functionally differentiate respect to the background field and
set all the fields to zero, we see that the background gluon
self-energy, written here in momentum space, is transverse
\begin{equation}\label{}
    k^\mu \tilde\Pi_{\mu\nu}(k)=0.
\end{equation}
If we differentiate \eqref{vm:WI_QED} respect to $\psi(y)$ and
$\bar\psi(z)$, and set all the field to zero, we get the Ward
identity of the gluon-quark vertex
\begin{equation}\label{vm:eq:WI_qqphot}
\partial_\mu \frac{\delta^3\tilde\Gamma}{\delta A^a_\mu(x)\delta\psi(y)\delta
    \bar\psi(z)}= -ig\bm T^a\left(\frac{\delta^2 \tilde\Gamma}{\delta\psi(x)\delta
    \bar\psi(z)} - \frac{\delta^2\tilde \Gamma}{\delta\bar\psi(x)\delta
    \psi(y)}\right).
\end{equation}
This relation is very important because it implies the well-known
relation (in QED) on the normalisation factors of the coupling
constant and the gauge field
\begin{equation}\label{vm:ZQED}
\boxed{Z_e = Z_A^{1/2}}
\end{equation}
With the help of \eqref{vm:ZQED}, we defined a renormalization group
invariant running coupling in QED, but also in QCD thanks to the
pinch technique. Finally, let us mention the relation between the
vertex and the self-energy
\begin{equation}\label{}
    q_1^\mu\tilde\Gamma^{abc}_{\mu\nu\rho}(q_1,q_2q_3) =
    -g\left(f^{abd}\tilde\Pi^{dc}_{\nu\rho}(-q_3)-f^{adc}\tilde\Pi^{bd}_{\nu\rho}(q_2)\right)
\end{equation}
easy derived from \eqref{vm:WI_QED} and showed by construction in
Sec.\ref{vm:sec:3gluons}.


\subsection*{Slavnov-Taylor identities in QCD}
We now derive the analogous relations for conventional QCD. They are
called Slavnov-Taylor identities or generalized Ward identities. We
will perform again a changement of variables but, this time, given
by the BRS transformations. Since the integrant of the generating
functional is invariant under these transformations, the Green's
functions of the theory also,
\begin{equation}\label{vm:STdef}
\boxed{\delta_{\text{BRS}}\langle0|T[A_\mu^a(x)\ldots]|0\rangle= 0}
\end{equation}
Where the dots stand for any fields. The Slavnov-Taylor identities
\eqref{vm:STdef} involve ghost fields. For instance, starting from
the trivial identity\footnote{In this section, I explicitly wrote
the color indices in fundamental representation. There are label
by the beginning of the Greek alphabet, $\alpha,\beta,\gamma$. }
\begin{equation}\label{}
\langle0|T[q_\alpha(x)\bar q_\beta(y)\phi^a(z)]|0\rangle=0
\end{equation}
and performing a BRS transformation, we arrive to the Slavnov-Taylor
identity
\begin{equation}\label{}
    -\frac{1}{q^2}q^\mu\Gamma_{\mu\alpha\beta}^b(q,p,r)\rouge{D^{-1}_{ba}(q)}
    = S^{-1}_{\alpha\gamma}(-p)\rouge{H^a_{\gamma\beta}(q,p,r)} -
    \rouge{\bar
    H^a_{\alpha\gamma}(-r,-q,-p)}S^{-1}_{\gamma\beta}(r),
\end{equation}
analogous to \eqref{vm:eq:WI_qqphot} but plagued with ghost
functions. $H$ is defined as follows,
\begin{equation*}\label{}
H^{b}_{\alpha\gamma}(q,p,r)S_{\gamma\beta}(r)D^{ba}(q) = -\bm
(T^a)_{\alpha\gamma}\int e^{-ipx}e^{-iry}
\langle0|T[q_\gamma(x)\phi^c(x) \bar q_\beta(y) \bar
\phi^a(0)]|0\rangle d^4xd^4y.
\end{equation*}
\begin{center}
\includegraphics[width=8cm]{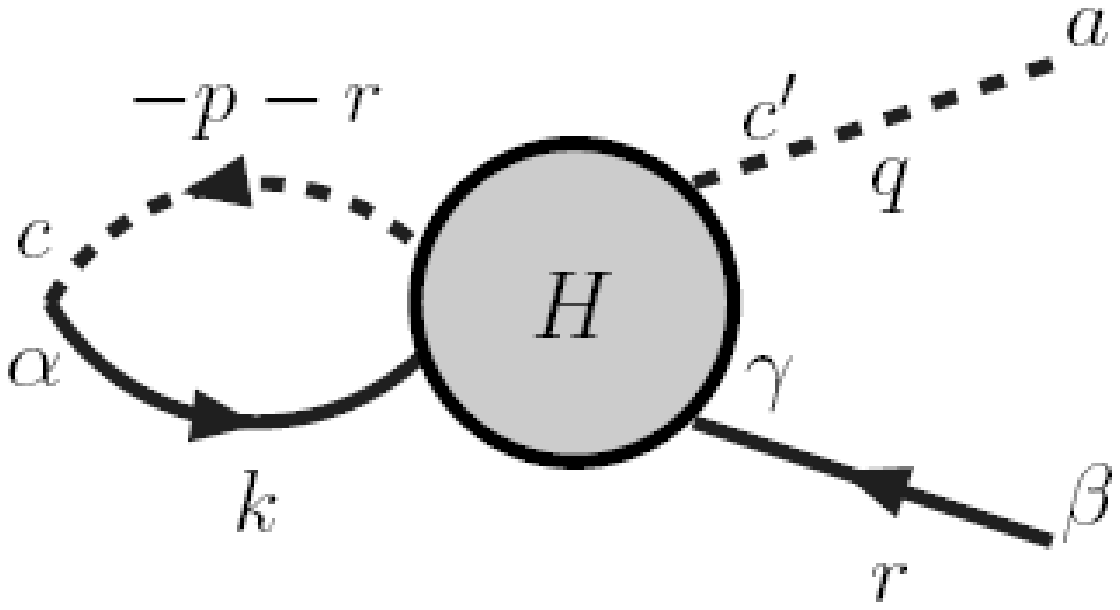}
\end{center}

We can also derive others relations using the BRS invariance of the
Green's functions such as
\begin{equation}\label{}
    q^\mu q^\nu D_{\mu\nu}(q) = q^\mu q^\nu D_{\mu\nu}^{(0)}(q),
\end{equation}
where $D_{\mu\nu}^{(0)}(q)$ is the free propagator of the gluon,
or
\begin{equation}\label{}
    p^\mu q^\nu k^\sigma\Gamma_{\mu\nu\sigma}(p,q,r) = 0.
\end{equation}
The first one indicates that to any order in perturbation theory
the longitudinal part of the propagator is equal to the
corresponding part of the free propagator. Note then that a
trivial gauge dependence remains in the full propagator.

\section{Dimensional Regularization}
To regularize divergent integrals we use the dimensional
regularization. This regularization scheme is useful for the pinch
technique because it preserves the gauge invariance of the theory.
Nevertheless, the rules of this scheme can lead to unconventional
formula such as
\begin{equation}\label{vm:DRrules}
    \int \frac{d^Dk}{(-k^2)^\alpha}=0\quad,\quad \alpha>0.
\end{equation}
We now proof this relation.

Applying a Wick rotation, the left-hand side of eq.
\eqref{vm:DRrules} may be written as
\begin{equation}\label{vm:app1}
\int \frac{d^Dk}{(-k^2)^\alpha} = i\frac{\pi^{D/2}}{\Gamma(D/2)}
\int_0^\infty (K^2)^{D/2-\alpha-1}dK^2,
\end{equation}
where $K^2=-k^2$. We see that Eq. \eqref{vm:app1} develops an
ultraviolet divergence for $D>2\alpha$ while it has an infrared
divergence for $D<2\alpha$. Thus the above integral has no
mathematically meaningful region in $D$. In order to give a
mathematical meaning to the integral in eq. \eqref{vm:app1}, we
split the integration in $K^2$ into the parts: The ultraviolet part
$K^2>\Lambda^2$ and the infrared part $K^2<\Lambda^2$,
\begin{equation}\label{}
\int \frac{d^Dk}{(-k^2)^\alpha} =i\frac{\pi^{D/2}}{\Gamma(D/2)}
[\int_0^{\Lambda^2} (K^2)^{D/2-\alpha-1}dK^2 +
\int^\infty_{\Lambda^2} (K^2)^{D/2-\alpha-1}dK^2].
\end{equation}
O the right-hand side of this equation, the first integral is
convergent for $D>2\alpha$ while the second one is convergent for
$D<2\alpha$. Here the space-time dimension $D$ acts as a regulator
for the infrared as well as ultraviolet. To distinguish the nature
of the divergences we designate $D=D_I$ for the first integral and
$D=D_U$ for the second. Performing the integration for $D_I>2\alpha$
and $D_U<2\alpha$ we obtain
\begin{equation}\label{vm:app2}
\frac{\Gamma(D/2)}{i\pi^{D/2}}\int \frac{d^Dk}{(-k^2)^\alpha} =
\frac{\Lambda^{D_I-2\alpha}}{\frac{1}{2}D_I-\alpha}-
\frac{\Lambda^{D_U-2\alpha}}{\frac{1}{2}D_U-\alpha}.
\end{equation}
We see that the two terms in Eq. \eqref{vm:app2} develop poles at
$D_I=D_U=2\alpha$ corresponding to the infrared and ultraviolet
divergence, respectively. The right-hand side of eq. \eqref{vm:app2}
can be continued analytically to arbitrary values of $D_I$ and $D_U$
and hence the constrains $D_I>2\alpha$ and $D_U<2\alpha$ can be
removed. If we identify $D_I$ with $D_U$ in Eq. \eqref{vm:app2}, the
right-hand side obviously vanishes.

\section{Feynman rules}

In this Appendix we list for completeness the Feynman rules in the
background field method in covariant gauges appearing in~\cite{Abb}.
Note that in this gauges, the Feynman rules for the quantum field
are the same that in the conventional formalism.
%

\noindent

\vspace{1cm}
\begin{tabular}{ll}                            
\begin{minipage}{2cm}
{\includegraphics[scale=0.5]{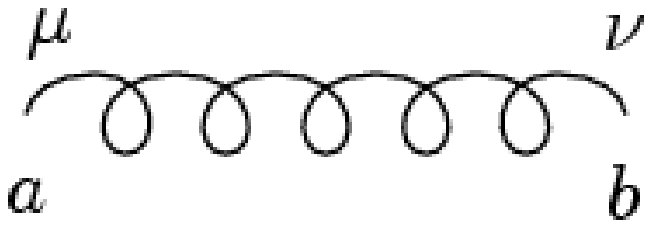}}
\end{minipage} &
\begin{minipage}{14.2cm}
\begin{equation}
\hspace{-2cm}
-i\left[g_{\mu\nu}-(1-\xi_Q)\frac{k_{\mu}k_{\nu}}{k^2}\right]\frac{\delta^{ab}}{k^2+i\epsilon}
\end{equation}
\end{minipage} \\
& \\
\end{tabular}

\begin{tabular}{ll}
\begin{minipage}{2cm}
{\includegraphics[scale=0.5]{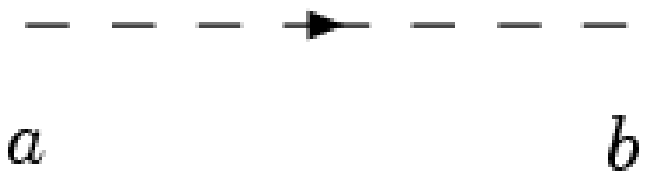}}
\end{minipage} &
\begin{minipage}{14.2cm}
\begin{equation}
\frac{i\delta^{ab}}{k^2+i\epsilon} 
\end{equation}
\end{minipage} \\
\end{tabular}
\vspace{1cm}
\begin{tabular}{ll}
\begin{minipage}{2cm}
{\includegraphics[scale=0.5]{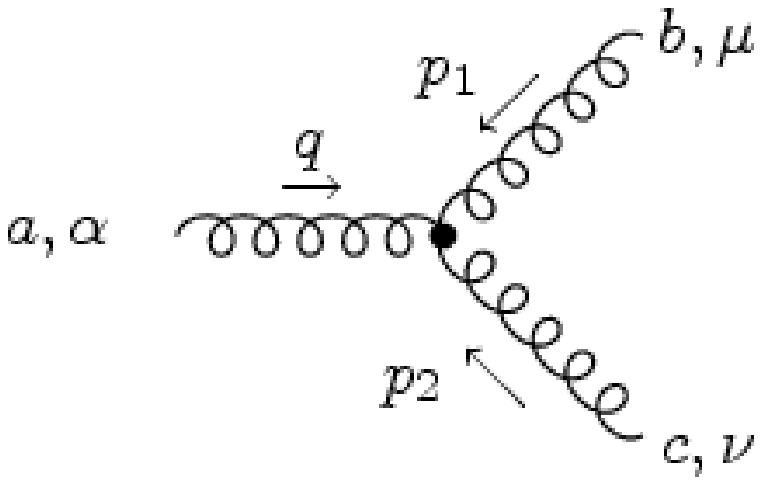}}
\end{minipage} &
\begin{minipage}{14.7cm}     
\begin{flushleft}
\begin{equation}
g f^{abc}\left[\left(q-p_1\right)_{\nu}g_{\mu\alpha}+
\left(p_2-q\right)_{\mu}g_{\nu\alpha} +
\left(p_1-p_2\right)_{\alpha}g_{\mu\nu}\right] \label{vm:3g_vertex}
\end{equation}
\end{flushleft}
\end{minipage}\\
\end{tabular}
\vspace{1cm} \noindent
\begin{tabular}{ll}
\begin{minipage}{2.2cm}
{\includegraphics[scale=0.4]{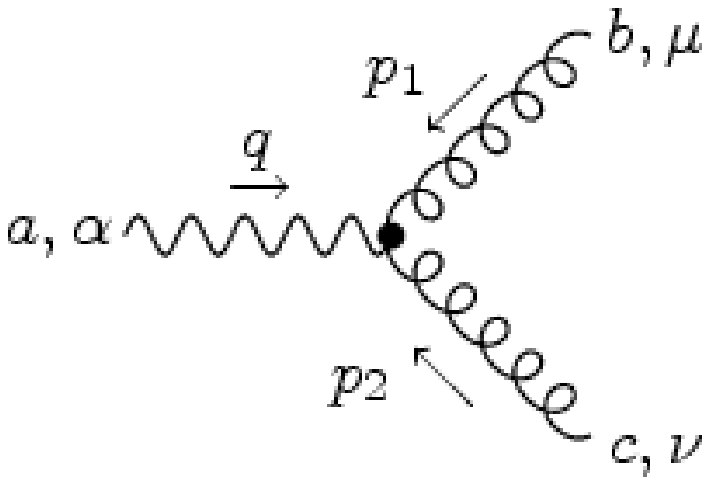}}
\end{minipage} &
\begin{minipage}{14.3cm}
\begin{flushright}
\begin{equation}
g
f^{abc}\left[\left(p_1-q+\frac{1}{\xi_Q}p_2\right)_{\nu}g_{\mu\alpha}+\left(q
-p_2-\frac{1}{\xi_Q}p_1\right)_{\mu}g_{\nu\alpha} +
\left(p_2-p_1\right)_{\alpha}g_{\mu\nu}\right] \!\!
\label{vm:bfm_3g}
\end{equation}
\end{flushright}
\end{minipage} \\
\end{tabular}
\vspace{1cm} \noindent
\begin{tabular}{ll}
\begin{minipage}{7cm}
{\includegraphics[scale=0.5]{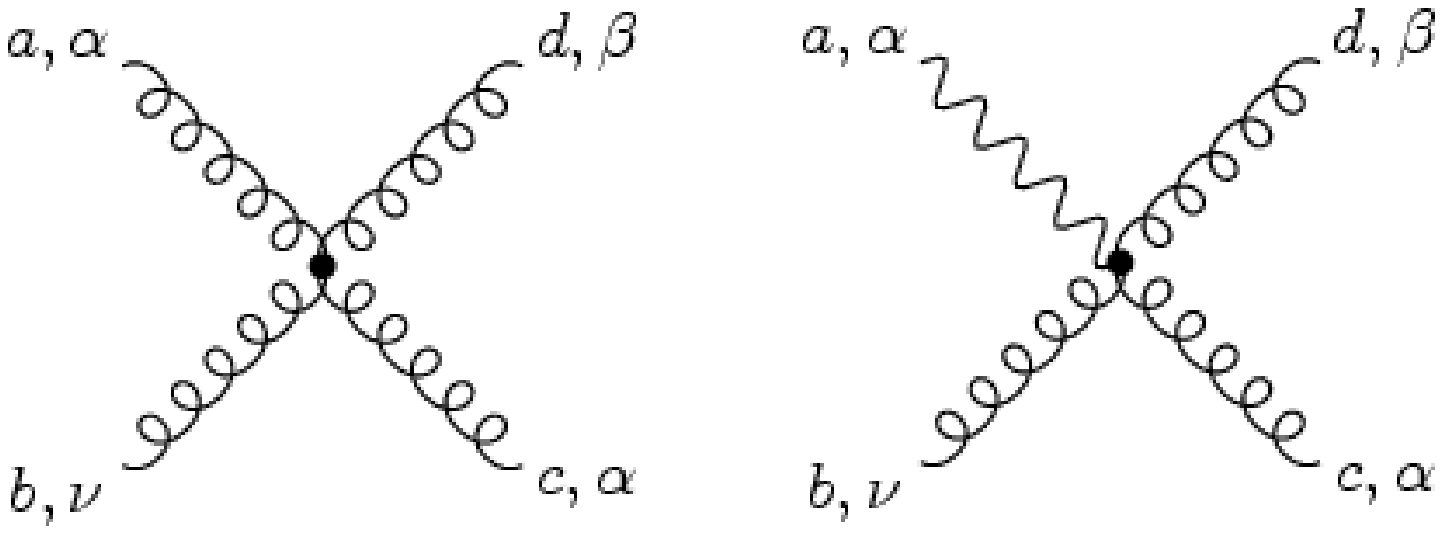}}
\end{minipage} &
\begin{minipage}{8.5cm}
\begin{flushright}
\begin{eqnarray}
&&-ig^2\bigg[ f^{abx}
f^{xcd}\left(g_{\mu\alpha}g_{\nu\beta}-g_{\mu\beta}g_{\nu\alpha}
\right) \bigg. \nonumber \\
&& \bigg. +  f^{adx} f^{xbc}\left(
g_{\mu\nu}g_{\alpha\beta}-g_{\mu\alpha}g_{\nu\beta}
 \right) \bigg.  \nonumber \\
&& + \bigg. f^{acx} f^{xbd}
\left(g_{\mu\nu}g_{\alpha\beta}-g_{\mu\beta}g_{\nu\alpha}\right)
\bigg] \label{vm:4gluon}
\end{eqnarray}
\end{flushright}
\end{minipage}  \\
\end{tabular}
\vspace{1cm} \noindent \\
\begin{tabular}{ll}
\begin{minipage}{2cm}
{\includegraphics[scale=0.5]{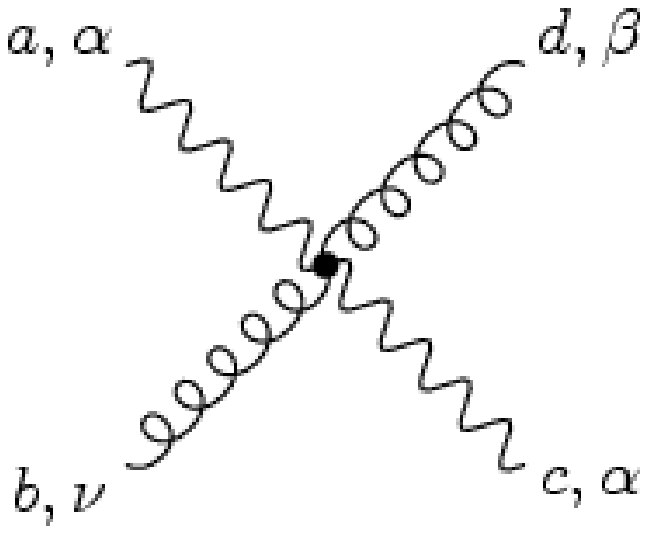}}
\end{minipage} &
\begin{minipage}{14.2cm}
\begin{flushright}
\begin{eqnarray}
&&-ig^2\bigg[ f^{abx}
f^{xcd}\left(g_{\mu\alpha}g_{\nu\beta}-g_{\mu\beta}g_{\nu\alpha}
+ \frac{1}{\xi_Q}g_{\mu\nu}g_{\alpha\beta}\right) \bigg. \nonumber \\
&& \bigg. +  f^{adx} f^{xbc}\left(
g_{\mu\nu}g_{\alpha\beta}-g_{\mu\alpha}g_{\nu\beta}
- \frac{1}{\xi_Q}g_{\mu\beta}g_{\nu\alpha} \right) \bigg.  \nonumber \\
 &&+ \bigg. f^{acx} f^{xbd}
\left(g_{\mu\nu}g_{\alpha\beta}-g_{\mu\beta}g_{\nu\alpha}\right)
\bigg] 
\end{eqnarray}
\end{flushright}
\end{minipage} \\
\end{tabular}
\vspace{1cm} \noindent
\begin{tabular}{ll}
\begin{minipage}{2cm}
{\includegraphics[scale=0.5]{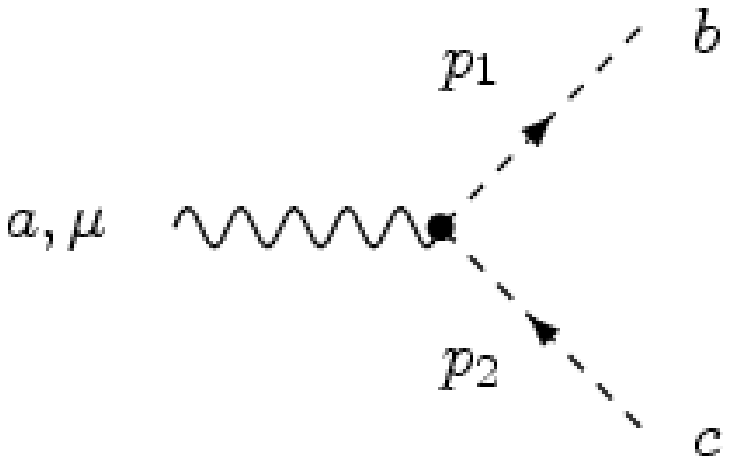}}
\end{minipage} &
\begin{minipage}{14.2cm}
\begin{flushright}
\begin{equation}
g f^{abc}(p_1+p_2)_{\mu} 
\end{equation}
\end{flushright}
\end{minipage}  \\
\end{tabular}
\vspace{1cm} \noindent
\begin{tabular}{ll}
\begin{minipage}{2cm}
{\includegraphics[scale=0.5]{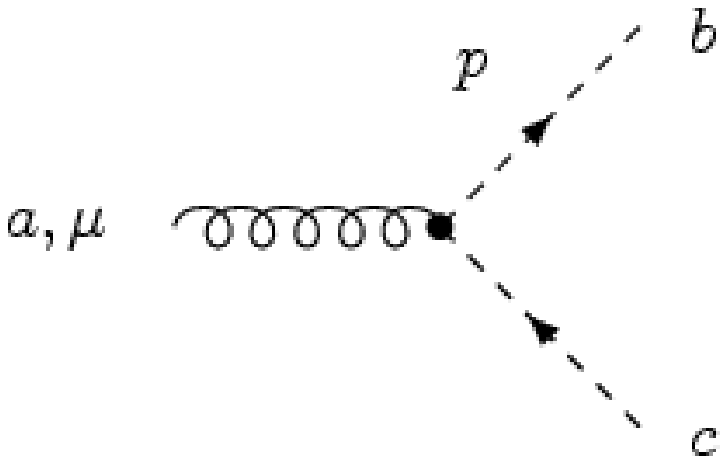}}
\end{minipage} &
\begin{minipage}{14.2cm}
\begin{flushright}
\begin{equation}
g f^{abc}p_{\mu} 
\end{equation}
\end{flushright}
\end{minipage}  \\
\end{tabular}
\vspace{1cm}
\begin{tabular}{ll}
\begin{minipage}{2cm}
{\includegraphics[scale=0.5]{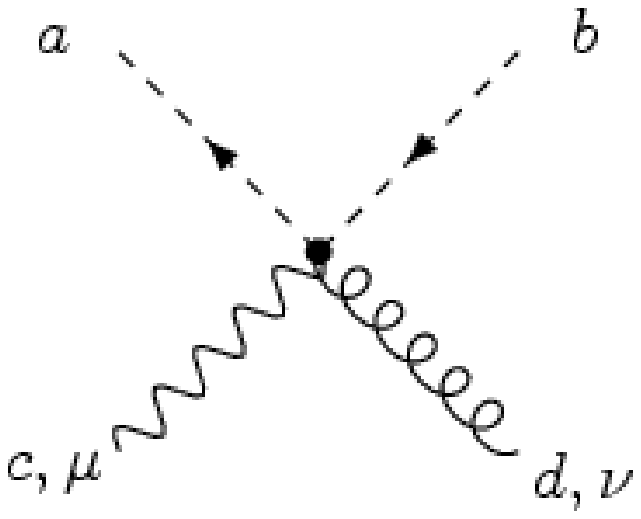}}
\end{minipage} &
\begin{minipage}{14.2cm}
\begin{flushright}
\begin{equation}
-ig^2 f^{acx}f^{xdb}g_{\mu\nu} 
\end{equation}
\end{flushright}
\end{minipage}  \\
\vspace{1cm}
\begin{minipage}{1cm}
{\includegraphics[scale=0.5]{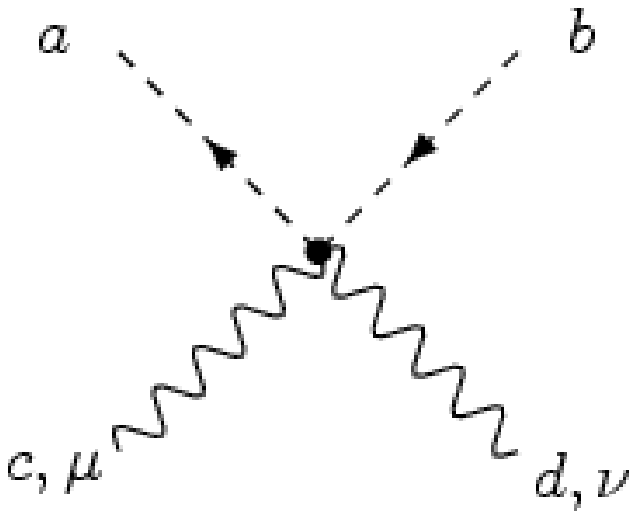}}
\end{minipage} &
\begin{minipage}{14.2cm}
\begin{flushright}
\begin{equation}
-ig^2 g_{\mu\nu}\left(f^{acx}f^{xdb}+f^{adx}f^{xcb}\right)
\end{equation}
\end{flushright}
\end{minipage}
\end{tabular}
\vspace{1cm}




\end{document}